\documentclass[preprint,12pt,authoryear]{elsarticle}

\usepackage{amssymb}
\usepackage{amsmath}

\usepackage{float}
\usepackage{verbatim}
\usepackage{hyperref}
\hypersetup{
    colorlinks=true,
    linkcolor=blue,   
    citecolor=blue,   
    urlcolor=blue,    
    filecolor=blue,   
    pdfborder={0 0 0} 
}

\usepackage{inputenc}

\usepackage{ulem}

\usepackage{changepage} 

\usepackage{multicol}
\usepackage{enumitem}
\usepackage{array}
\usepackage{longtable}

\usepackage{multirow}%

\usepackage[breakable]{tcolorbox}
\newtcolorbox{shadedbox}{
  colback=gray!10,
  colframe=white,
  breakable,
  left=1pt,
  right=1pt,
  top=1pt,
  bottom=1pt,
  overlay={}
}

\usepackage[authoryear]{natbib}

\journal{xx}

\begin{document}

\begin{frontmatter}



\title{Interpersonal Trust Among Students in Virtual Learning Environments: A Comprehensive Review} 


\author[label1,label2]{Marcelo Pereira Barbosa}
\ead{marcelo.pereira@ifpi.edu.br}
\author[label2]{Rita Suzana Pitangueira Maciel}
\ead{rita.suzana@ufba.br}
\affiliation[label1]{organization={Instituto Federal do Piauí},
             addressline={},
             city={São João do Piauí},
             postcode={},
             state={Piauí},
             country={Brazil}}

\affiliation[label2]{organization={Universidade Federal da Bahia},
             addressline={},
             city={Salvador},
             postcode={},
             state={Bahia},
             country={Brazil}}

\begin{abstract}
Interpersonal trust is recognized as one of the pillars of collaboration and successful learning among students in virtual learning environments (VLEs). This systematic mapping study investigates attributes, phases, and features that support interpersonal trust among students in VLEs. Analyzing 46 articles, we identified 37 attributes that influence phases of acquiring and losing trust, categorized into four themes: Ability, Integrity, Affinity, and Non-Personal Factors. Attributes such as collaborative and ethical behavior, academic skills, and higher grades are often used to select peers, mainly through recommendation systems and user profiles. To organize our findings, we elaborated two conceptual maps describing the main characteristics of trust definitions and the attributes classification by phases and themes.
\end{abstract}





\end{frontmatter}

\section{Introduction}\label{}

In daily life, trust is usually perceived as an individual's willingness to accept vulnerability based on positive expectations of others' intentions or actions in situations involving interdependence and risk \citep{mahamud2021predicting}. Trust sustains human relationships as if without minimal trust, it is almost impossible to establish and maintain relationships over a long period of time \citep{bachmann2001trust}. Moreover, it is strongly related to psychological contracts, based on mutual expectations and reciprocal faith that each party to a relationship has that the other party will keep their promises \citep{snyman2021framework}.

Trust operates within the cognitive and psychological domain as a motive for behavior at the interpersonal level to shape social exchanges as a multidimensional and dynamic phenomenon, constituted by a life cycle, which can be acquiring, maintain, losing and restored \citep{kutsyuruba2015lifecycle}. Trust will evolve and change over time in relationships, as knowledge and information about the trustworthiness of other parties also evolve in these relationships \citep{fachrunnisa2010state}. A negative signal can destroy trust, and the cost of restoring it is very high.

Trust relationships can manifest among individuals (interpersonal) and between individuals and institutions (organizational). Interpersonal trust constitutes an expectation held by an individual that another individual or group can be trusted, while organizational trust is based on the circumstances of a robust and coherent institutional structure, in which trust is generated on an institutional basis, i.e., in the form of trust in the system \citep{bachmann2001trust}.

As remote learning continues to grow in popularity, online courses are evolving beyond content delivery to foster collaboration, communication, and a sense of community among students and educators \citep{pham_2020_the_role_of}. As a result, it has become imperative that virtual learning environments (VLEs) address the need for trust-enhancing features that can significantly improve the learning experience. In online courses,  trust relationships affect the intention to share knowledge, reduce dropout rates, and leverage student academic success \citep{younas_2021_role_of_design} \citep{sedighehm_2018_effect}
\citep{ghosh2001student} 
\citep{horvat2013students}. Students who seldom interact with their peers experience more negative emotions and exhibit diminished effectiveness in their learning process. In this context, where face-to-face interactions are limited, acquiring and maintaining trust is essential to prevent students from feeling isolated and disengaged. However, promoting interpersonal trust among students within VLEs is not trivial \citep{wang_2012_emotions_and_pair}.

Over the years, several research papers have addressed trust in E-learning courses, focusing mainly on the relationships between learners and resources, i.e. the relationship between learners and VLE resources \citep{younas_2021_role_of_design, pham_2020_the_role_of, wu_2015_trust_evalueation}, but lacking intrinsic aspects of social relationships and the factors that influence interpersonal trust (mutual trust relationship between students) \citep{zhang_2022_a_novel_precise}. Regarding secondary studies, some address trust among students such as:  analyzed it as a precondition for collaboration in VLEs \citep{lohikoski2014virtual}, trust in VLE platforms \citep{medema_2014_multi_loop} and trust in online assessment systems \citep{nigam_2021_a_systematic_review}. These secondary studies contribute to the state of the art of trust but do not fill the gaps in the literature regarding interpersonal trust between students and the factors that support interpersonal trust in VLEs. A lack of focused research on interpersonal trust in virtual learning environments can hinder the development of VLEs that prioritize this crucial aspect. Ensuring interpersonal trust is essential to foster a positive learning experience \citep{khan2023modelling}. 

Likewise, given the lack of studies in the literature on the factors influencing interpersonal trust among students in VLEs, we propose conducting this Systematic Mapping Study (SMS). Our main objective is to  systematically identify the elements influencing trust among students in this context. This investigation aims to address the following research question (RQ):

\begin{adjustwidth}{1cm}{1cm} 
\centering
\textbf{\textit{What Characteristics influence interpersonal trust among students in Virtual Learning Environments?\\}}
\end{adjustwidth}

We performed this SMS following the protocol proposed by \cite{kitchenham2015evidence}, which included the objectives, research questions, and selection criteria (inclusion and exclusion) for the articles, and, in the end, 46 articles were accepted. The data was processed using thematic analysis techniques \citep{clarke2013successful,diba2011recommended}. This gave us an in-depth understanding of the data collected and insight into the key attributes that influence interpersonal trust among student VLEs. In summary, this article yields six main contributions: 

\begin{itemize}

\item We offer an updated, original, and comprehensive panorama of trust aspects addressed in the e-learning domain by 46 primary studies. To assist readers, we summarize the key takeaways of each research question. We have made our data set available for replication studies or other related research works \hypertarget{DATA}{(\href{https://bit.ly/4cTJ2gc}{https://bit.ly/4cTJ2gc})}.

\item We collected 23 trust definitions used in the literature and organized key terms into a conceptual map (Figure \ref{fig:definition_trust}). Moreover, we elaborated on a definition of trust specifically for online courses.

\item We identified 37 attributes that influence interpersonal trust among students, organized them into four themes and sixteen categories and classified them according to their influence on the trust phases. For better understanding, elaborate a mind map (Figure \ref{fig:categories_interpersonal_trust}). As a highlight, we proposed two categories to group attributes within the e-learning domain.

\item We identified trust phases that are addressed and those that are still not being covered, organizing and classifying them by their association with trust phases (as mentioned earlier).

\item We found that seven features are recurrently related to supporting trust in VLEs and associated them with the trust phases and attributes.

\item We propose seven potential research opportunities that should be pursued based on our findings.\\
\end{itemize}

The results of this study are intended to contribute to teachers, students, software engineers, and researchers:
\begin{itemize}

\item For teachers, they can use the information in this SMS to develop methodological strategies that foster student trust based on understanding the factors that influence it. Teaching methods can be adapted to focus on developing trust and, as a result, improve collaboration between students. 

\item For students, this SMS helps identify the characteristics that increase and decrease trust, allowing them to develop and improve their soft and hard skills to demonstrate greater trustworthiness. 

\item For software engineers interested in online courses can use the data from this SMS to develop features that support trust among students in virtual learning environments. The classification of interpersonal attributes is valuable in this process because it allows developers to identify and incorporate attributes that increase trust and mitigate those that may decrease trust. 

\item For researchers, exploring these phases and attributes is essential to developing strategies that enhance trust, mitigate trust erosion, and ultimately foster a more collaborative and effective virtual learning environment where trust significantly enhances student collaboration. 

\end{itemize}

This article is structured as follows: Background and Related work,  we present the theoretical basis on trust and secondary studies that have some similarity to ours; Research methodology,  we show the methodology and the steps used to perform the SMS; Results, we present the findings of our research; Discussion, where we perform an analysis of the results; Threats to validity,  we discuss the threats to our study; and finally, Final considerations and future work,  we summarize what has been obtained from our research.

\section{Background}

Since the seminal study by \cite{deutsch1958trust}, research on trust has been performed in many fields, including sociology \cite{sztompka1999trust}, psychology \citep{robinson1996trust}, economics \citep{knack2003building, berggren2006free}, communication \citep{cheung2013interweaving}, leadership \citep{burke2007trust}, and the implementation of self-managed work teams \citep{trainer2018bridging}.

Over the years, trust has been defined in various ways, encompassing the willingness of individuals within a social unit to rely on others \citep{rotter1967new}, the willingness of one party to be vulnerable to another's actions based on the expectation that the latter will perform actions necessary to the former regardless of the ability to monitor or control them \citep{mayer1995integrative}, the readiness to act based on another person's words, actions, and decisions, and the psychological state involving the intention to accept vulnerability based on positive expectations of another's intentions or behavior\citep {rousseau1998not}.

 Interpersonal trust was initially presented in the literature as an expectation held by an individual or group that a word, promise, verbal or written statement of another individual or group can be relied upon \citep{rotter1967new}. \cite{borum2010science} argues that interpersonal trust operates under conditions of recognized interdependence and is characterized by a willingness to accept vulnerability and risk based on confident expectations that another person's future actions will produce good results. Interpersonal trust is a relationship that involves at least two people: (i) \textit{trustor} is the agent who plays the role of depositing the trust, and (ii) \textit{trustee} is responsible for managing and fulfilling that trust.
This means that the trust that the trustor has in the trustee corresponds to the belief that the trustee's future actions will lead to positive results \citep{elghomary_2019_a_comparative_analysis}. Moreover, interpersonal trust is based on identifiable attributes that another person is expected to possess and is a multidimensional construct that includes cognitive and emotional dimensions \citep{schaubroeck2013developing}.

\cite{mayer1995integrative} argues that the factors that influence trust are related to three characteristics of the trustee: ability, integrity, and benevolence. Ability is the group of skills, competencies, and characteristics that allow a party to influence some specific domain. The ability domain is specific because the trustee may be highly competent in some technical area, which gives him trust in the tasks related to that area. However, the trustee exercising the trust may have little aptitude, training, or experience in academic areas, e.g., in interpersonal communication.  Integrity involves the trustor perception that the trustee adheres to a set of principles that he considers acceptable. However, if this set of principles is not considered acceptable, it will not be deemed to have integrity for the trustor. Issues such as consistency of past actions, sense of justice, and character are examples of factors that make up the integrity of the trustee. Finally, benevolence is the degree to which a trustee is believed to want to do good to the trustor, based on the trustee's altruism and going beyond a motive to gain some advantage.

Regarding how trust can be established, trust can develop beyond the direct relationship between trustor and trustee, with the help of a third party, forming a network of trust \citep{mcevily2021network} and \cite{jones2021tangled}. In this case, trust occurs between individuals who are not necessarily directly connected. A network of trust is the idea that trust is not only formed through direct interaction but also indirect connections between individuals. The trust network has two distinct forms: second-hand trust and prototrust \citep{mcevily2021network}. 

Second-hand trust refers to the partial overflow of relational trust to socially close and indirectly connected actors in a strong and reciprocal relationship sustained by a common third party, in which two of the actors are not directly connected to each other but are connected to a third party with reciprocal bonds of trust. Prototrust differs from second-hand trust in that prototrust occurs mainly in affiliation networks, which involve joint participation or membership in collectivizes, such as social groups, clubs, and professional associations \citep{mcevily2021network}.

Moreover,  interpersonal trust has some properties \citep{elghomary_2019_a_comparative_analysis}:
 \textit{Asymmetrical}, not reciprocal or equivalent between two entities, it is  personalized and subjective since two people have different opinions about the same person;
\textit{Transitive}, considering that A trusts B, who trusts C, then A tends to trust C; 
\textit{Propagative}, given a social environment, trust information can pass from one entity to another, allows trust values to be aggregated between any two entities that are not directly connected; and \textit{Dynamic}, changes occur according to the context, over time, or for a specific task or objective. The dynamics of trust involve the evolutionary phases: Acquiring, losing, Maintain, and Restore \citep{currall2003fragility, fachrunnisa2010state}.

Trust is dynamic, and the level of trust in a relationship progresses, starting close to zero due to a lack of information about the counterpart and growing through trust-building actions until the maintenance phase when it remains stable with minimal fluctuations. Once trust has been established, demands for evidence decrease, but continuous information is key to maintaining it \citep{currall2003fragility, fachrunnisa2010state}. When built, it still requires new evidence of trustworthiness to be maintained. However, it can quickly be eroded if solid evidence of a lack of trust exists. The process of repairing trust is not easy, and, in some circumstances, no repair initiative will be able to restore it \citep{kahkonen2021employee}. With regard to the dynamics and changing phases of trust, the literature points out that building trust is difficult to establish in online environments \citep{anwar_2021_supporting_privacy}.


\section{Related Work}

This section presents the works related to our SMS, i.e., secondary studies that addressed trust in VLEs.

To identify the related works, we elaborate a search string built considering three domains. The first domain is related to our research problem and comprises terms related to interpersonal trust. The second domain includes terms relating to our area of interest, virtual learning environments. The third domain comprises terms capable of identifying secondary studies according to \cite{napoleao2021establishing}. We limited our research to the period from January 2012 to October 2023.

\begin{adjustwidth}{0.3cm}{0.3cm}
\begin{shadedbox} 
\textbf{Domain 1:}
(``interpersonal trust" OR trust OR trustworthiness OR trusting OR trustworthy OR trusted)
\end{shadedbox}
\end{adjustwidth}

\begin{adjustwidth}{0.3cm}{0.3cm}

\begin{shadedbox} 
\textbf{Domain 2:}\\
``e-learning" OR ``online-courses" OR  ``online courses" OR   ``online learning" OR   ``virtual learning" OR ``distance learning" OR ``learning platforms" OR ``Learning Management Systems" OR ``Computer-Supported Collaborative Learning" OR ``Computer Supported Collaborative Learning" OR ``CSCL" OR  ``VLE" OR ``Virtual Learning Environment" OR mooc OR  moocs)
\end{shadedbox}
\end{adjustwidth}

\begin{adjustwidth}{0.3cm}{0.3cm}
\begin{shadedbox} 
\textbf{Domain 3:} \\
(``literature review"  OR  ``systematic mapping"  OR  ``systematic review"  OR  ``mapping study"  OR  ``systematic map").
\end{shadedbox}
\end{adjustwidth}

\begin{adjustwidth}{0.3cm}{0.3cm}

\begin{shadedbox} 
\textbf{Search string }\\
(``interpersonal trust" OR trust OR trustworthiness OR trusting OR trustworthy OR trusted)  AND  (``e-learning" OR ``online-courses" OR  ``online courses” OR   ``online learning" OR   ``virtual learning" OR ``distance learning" OR ``learning platforms" OR ``Learning Management Systems" OR ``Computer-Supported Collaborative Learning" OR ``Computer Supported Collaborative Learning" OR ``CSCL" OR  ``VLE" OR ``Virtual Learning Environment" OR mooc OR  moocs)  AND ( ``literature review"  OR  ``systematic mapping"  OR  ``systematic review"  OR  ``mapping study"  OR  ``systematic map").
\end{shadedbox}
\end{adjustwidth}

We applied the string into libraries IEEE Xplorer, ACM Digital Library, and Scopus, returning 22 studies. Initially, we analyzed the titles and abstracts of the 22 studies and excluded those not aligned with the topic under study. Although none of the secondary studies are strictly linked to ours, i.e., do not share similar objectives, research questions, or results, we selected the four closest ones for a deeper analysis. Table \ref{tab:Related_Work} shows the selected studies.

\small

\begin{table}[!ht]
\centering
\renewcommand{\arraystretch}{1.1} 
\begin{tabular}{p{0.6cm} p{6.5cm} p{1.5cm} p{2.cm} p{1cm}} 
  \hline
  \textbf{Cod} & \textbf{Title} & \textbf{Authors} & \textbf{Source} & \textbf{Year} \\ \hline
  \hypertarget{S1}{S1} & Virtual collaboration competence requirements for entrepreneurship education in sparsely populated areas & \cite{lohikoski2014virtual} & Scopus & 2014 \\
  
  \hypertarget{S2}{S2} & Multi-Loop Social Learning for Sustainable Land and Water Governance: Towards a Research Agenda on the Potential of Virtual Learning Platforms & \cite{medema_2014_multi_loop} & Scopus & 2014 \\
   
  \hypertarget{S3}{S3} & A Systematic Review on AI-based Proctoring Systems: Past, Present and Future & \cite{nigam_2021_a_systematic_review} & Scopus & 2021 \\
   
  \hypertarget{S4}{S4} & Academic dishonesty and trustworthy assessment in online learning: A systematic literature review & \cite{surahman_2022_academic_dishonesty} & Scopus & 2022 \\ \hline
\end{tabular}
\caption{Related work}
\label{tab:Related_Work}
\end{table}

\normalsize

\hyperlink{S1}{S1} analyzed the preconditions for collaboration in VLEs and recommended face-to-face interaction at the beginning of projects. Face-to-face interaction is an antecedent in the peer trust construct, while the intention to collaborate increases when there is peer trust in VLE. Results point that face-to-face interaction is one of the motivators for building trust. 

\hyperlink{S2}{S2} provided insight into the conditions and possibilities that lead to multi-loop social learning on  VLE platforms for building trust, co-producing knowledge, creating meaning, critical self-reflection, vertical and horizontal collaboration, and conflict resolution. Although \hyperlink{S2}{S2} study analysis trust building in VLEs, it lacks to analyze other phases of trust: maintain, losing, and restore trust.

\hyperlink{S3}{S3} performed a systematic literature review on AI-based online assessment supervision systems. The study considers trust to be important in online assessments, both for the appraiser and the appraiser, and leaves an open question on how trust-based online assessment monitoring systems should be built. 

\hyperlink{S4}{S4} investigated the factors contributing to academic dishonesty and trustworthy assessment in e-learning. The results indicate academic dishonesty among students in VLEs, which includes plagiarism, cheating, collusion, and other dishonest behavior. Academic dishonesty is strongly associated with individual factors such as laziness and lack of ability and situational factors such as peer influence, course pressure, and ease of access to information. Literature suggests that online supervision systems and anti-plagiarism tools can improve trust in assessment and reduce academic dishonesty. 

Previous studies have analyzed trust in VLEs but have focused on technological aspects, with little consideration of interpersonal trust between students. In contrast, our study identifies factors that influence students' interpersonal trust and explores the attributes and phases of trust in VLEs. To this end, we conducted our SMS.


\section{Research Methodology}
\label{sec:research}
    
SMS provides a broad overview of a research area to obtain and analyze state-of-the-art on a topic \citep{kitchenham2011using}. 
Towards systematizing the development of this study and answering our research questions, we conducted an SMS following the protocol proposed by \cite{kitchenham2007guidelines} and \cite{kitchenham2015evidence}.

Conducting the review means putting the protocol into practice, which comprises three main phases: (i) \textit{Planning}, define the SMS protocol comprising research questions,  specifying the search strings and strategies for selection of primary studies according to inclusion and exclusion criteria; 
(ii) \textit{Conduction}, involves the specification and application of the search string in chosen databases, selection of primary studies through the application of inclusion and exclusion criteria, as well as the extraction and synthesis of data from these studies,
and (iii) \textit{Report of results}, involves documenting all the steps and results of the research questions. In section \ref{sub:planning},  we detail the steps performed out.

\subsection{Planning}
\label{sub:planning}

SMS primary goal is to identify the motivating factors that influence interpersonal trust among students in  VLEs through a comprehensive literature analysis. The main elements of an SMS are the study objectives, research questions, search string, selection criteria, data extraction, and synthesis of the data extraction. Concerning the goal, we used GQM (Goal Question Metric) to define it \citep{solingengoal}.

\begin{shadedbox}
    
The goal is to analyze primary studies of Interpersonal Trust in  VLE for the purpose of identifying the panorama with respect to
trust definition, attributes, phases, variation, and solutions from the point of view of researchers in the
context of academic literature. 
\end {shadedbox}

We developed this SMS planning with the aim of thoroughly conducting the protocol to answer the following main research question:\\

\textbf{What Characteristics influence interpersonal trust among students in Virtual Learning Environments?}\\

Therefore, to better achieve our goal, we divided the main RQ into two RQs, as listed below:

\begin{adjustwidth}{0.5cm}{0cm}	
\textbf{\textit{RQ 1. What trust characteristics are considered in virtual learning environment?}}
\end{adjustwidth}
    
\begin{adjustwidth}{0.9cm}{0cm}
   \textit{ RQ 1.1 How is trust defined?}
\end{adjustwidth}

     \begin{adjustwidth}{1.2cm}{0cm}
          
  \textit{Rationale}: By analyzing the definitions, we aim to identify the fundamental conceptual bases of trust and understand its essential characteristics.  This analysis will allow us to clarify conceptual nuances and identify convergences or divergences that influence the understanding of trust.\\
       
\end{adjustwidth}
    
 \begin{adjustwidth}{0.9cm}{0cm}	
    \textit{RQ 1.2 What interpersonal trust attributes are considered in Virtual Learning Environments?}	
\end{adjustwidth}

 \begin{adjustwidth}{1.2cm}{0cm}	
   \textit{Rationale} - We intend to identify attributes, which are specific characteristics, measurable or identifiable properties of an entity. Therefore, we intend to collect data and group it so that it can be used to build collaborative VLEs with the aim of supporting interpersonal trust among students.\\
\end{adjustwidth}

\begin{adjustwidth}{0.9cm}{0cm}	
\textit{RQ 1.3 In which evolutionary phases of trust can students' interpersonal trust attributes be classified?}
 \end{adjustwidth}

\begin{adjustwidth}{1.2cm}{0cm}	

\textit{Rationale} - We intend to identify the trust phases that are supported on VLEs and  which attributes are related to trust specific phases: acquiring, losing, maintain, and restore trust.\\
 \end{adjustwidth}

\begin{adjustwidth}{0.5cm}{0cm}
\textit{\textbf{RQ 2. Which functionalities in VLEs  support student interpersonal trust?}}
 \end{adjustwidth}

\begin{adjustwidth}{1.2cm}{0cm}	
\textit{Rationale} - We want to identify proposed features that support interpersonal trust among students, what trust aspects are supported and what needs better attention in VLE.\\
 \end{adjustwidth}

To assess the completeness and accuracy of the search string, we used various articles to control its effectiveness. These articles are primary studies on interpersonal trust in e-learning environments that we became aware of from an ad hoc search of the literature and which should be included in our SMS (See table \ref{tab:Control_Studies}). Therefore, to find research works of interest, we defined a search string based on terms extracted from known articles, and which should appear in the set of articles that the string should return. After several calibrations, the following search string returned all the control articles \citep{kitchenham2015evidence}.

To construct the search string to identify primary studies, we used two domains. The first domain is related to our research problem and is made up of terms related to interpersonal trust. The second domain is made up of terms relating to our area of interest, virtual learning environments. For each domain, we used keywords that allowed us to broaden our search and identify all the studies relevant to our research.

\begin{adjustwidth}{0.3cm}{0.3cm}
 
\begin{shadedbox}
\textbf{Domain 1:}  \\
\textit{``interpersonal trust'' OR trust OR trustworthiness OR trusting OR trustworthy OR trusted}

\end{shadedbox}
\end{adjustwidth}

\begin{adjustwidth}{0.3cm}{0.3cm}

\begin{shadedbox} 
\textbf{Domain 2:}\\
``e-learning" OR ``online-courses" OR  ``online courses" OR   ``online learning" OR   ``virtual learning" OR ``distance learning" OR ``learning platforms" OR ``Learning Management Systems" OR ``Computer-Supported Collaborative Learning" OR ``Computer Supported Collaborative Learning" OR ``CSCL" OR  ``VLE" OR ``Virtual Learning Environment" OR mooc OR  moocs)
\end{shadedbox}
\end{adjustwidth}

\begin{adjustwidth}{0.3cm}{0.3cm}
 
\begin{shadedbox}

\textbf{Search string:} \\
\textit{(``interpersonal trust" OR trust OR trustworthiness OR trusting OR trustworthy OR trusted) AND (``e-learning" OR ``online-courses" OR  ``online courses” OR   ``online learning" OR   ``virtual learning" OR ``distance learning" OR ``learning platforms" OR ``Learning Management Systems" OR ``Computer-Supported Collaborative Learning" OR ``Computer Supported Collaborative Learning" OR ``CSCL" OR  ``VLE" OR ``Virtual Learning Environment" OR mooc OR  moocs)}
\end{shadedbox}
\end{adjustwidth}

\begin{table}[!ht]
\begin{tabular}{p{10cm}ll}
\hline
\textbf{Title} & \textbf{Source} & \textbf{Cod} \\
\hline
A Comparative Analysis of OSN and SIoT Trust Models for a Trust Model Adapted to MOOCs Platforms & ACM & \hyperlink{A28}{A28} \\
A collective intelligence approach for building student's trustworthiness profile in online learning & Scopus & \hyperlink{A7}{A7} \\
Case Study of Collaborative Learning in a Massive Open Online Course & IEEE & \hyperlink{A33}{A33} \\
Dynamic Peer Recommendation System based on Trust Model for sustainable social tutoring in MOOCs & IEEE & \hyperlink{A30}{A30} \\
In MOOCs we trust: Learner perceptions of MOOC quality via trust and credibility & Scopus & \hyperlink{A26}{A26} \\
Trust in pandemic-induced online learning: Competitive advantage of closure and reputation & Scopus & \hyperlink{A39}{A39} \\
\hline
\end{tabular}
\caption{Articles used to control the Search String precision}
\label{tab:Control_Studies}
\end{table}
 To guide the select the studies, we defined three inclusion criteria (IC) and eight exclusion criteria (EC) (Table  \ref{tab:criteria} ).

\begin{table} [!ht]
\centering
\begin{tabular}{lp{11cm}}
\hline
\textbf{COD} & \textbf{Inclusion (IC) and Exclusion Criteria (EC)} \\
\hline
IC1 & Studies address interpersonal trust issues \\
IC2 & Study is peer-reviewed \\ 
IC3 & Study is written in English \\
EC1 & Studies before 2012\\
EC2 & Duplicate studies \\
EC3 & Not peer-reviewed \\
EC4 & Study not in English \\
EC5 & Complete version is not available \\
EC6 & Study is an old version \\
EC7 & Study is a not primary one \\
EC8 & Study that no address interpersonal trust \\
\hline
\end{tabular}
\caption{Selection Criteria}
\label{tab:criteria}
\end{table}

In this SMS, we did not perform a quality assessment of the articles, as we wanted to consider as many articles as possible for our study. As we considered both theoretical and practical aspects in our analysis, the quality assessment could limit the set of articles. Quality assessment is essential in systematic reviews to determine the rigor and relevance of primary studies; however, no quality assessment needs to be performed in SMS so that as many articles as possible can be entered \citep{petersen2015guidelines, kitchenham2015evidence}.

We used the publishers digital libraries IEEE Xplorer\footnote{\href{https://ieeexplore.ieee.org}{https://ieeexplore.ieee.org}}, ACM Digital Library\footnote{\href{https://dl.acm.org}{https://dl.acm.org}}, and the Scopus\footnote{\href{https://www.scopus.com}{https://www.scopus.com}} indexer to identify the articles Because the publisher's digital libraries only return articles from the publisher, we considered inserting an indexer so that we could have a wider reach when searching for articles. Initially, the period selected was from January 2012 to December 2023.  After completing the analysis and writing of this SMS, we performed a new search in the databases from January 2024 to June 2024, with the aim of identifying new articles of interest to our research. However, we didn't identify any articles relevant to our research.

\subsection{Conducting}

\begin{figure}[!ht]
\includegraphics[width=\textwidth]{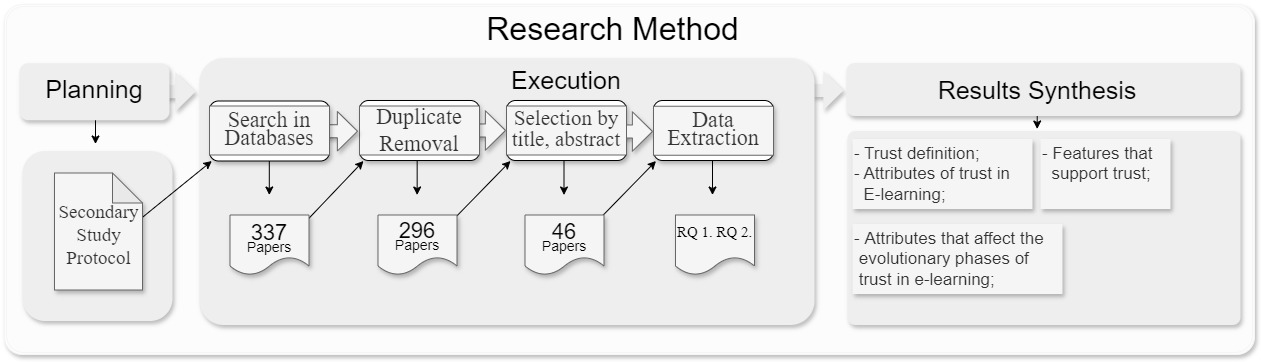}
\caption{Research Methodology}
\label{fig:research_methodology}
\end{figure}

The SMS was conducted in strict accordance with the planning proposed in the protocol, using Parsifal\footnote{\href{https://parsif.al/}{https://parsif.al/}} as a supporting tool. The methodological process is shown in the figure \ref{fig:research_methodology}. We searched for primary studies in the ACM, IEEE, and Scopus databases. The search string was configured for each search engine databases specificity. We applied the terms of the string to the title of the publication and the abstract and applied the filter to studies published between January 2012 and December 2023. 

From the Search String, we were able to identify a total of 337 articles. Of these, 146 were located in the IEEE database, of which 15 were considered suitable after careful analysis. In the ACM, our search revealed 39 articles, of which 4 were selected for inclusion. In addition, by exploring the Scopus indexer, we found 152 articles, 27 of which passed the selection process.

In the first step of the analysis, we checked for duplicate articles. This process was performed automatically on the Parsifal platform, and 41 duplicate articles were removed. In the second step, we performed the selection process and applied the inclusion and exclusion criteria, evaluating each article by titles and abstracts. This step was performed by two researchers (the first author and a master's student from our research group) and validated by a third (the second author). Each researcher applied the selection criteria individually, and then the results were compared. 
The disagreements were discussed and resolved as per \cite{kitchenham2015evidence} guidance.
For those articles where it was not possible to apply the inclusion and exclusion criteria from the title and abstract, the article was read in full to determine its inclusion/exclusion in the SMS. 

The process resulted in 46 studies being included and 291 removed. To ensure that the process of applying the inclusion and exclusion criteria was performed effectively, a third researcher from our research group applied the inclusion and exclusion criteria to 10\% of the excluded articles. There was no divergence from the results previously obtained. The final step was to read the full text and extract the data from each study. At the end of our analysis, we performed an update process, using the search string in the databases to identify articles up to June 2024. We identified 10 articles in this process, but none of them met the selection criteria.

The data extraction process was performed using a data extraction form, which was designed to collect all the information needed to answer the research questions. We began by performing a testing process of the data form, with a pilot extraction in some primary studies used as control studies \citep{kitchenham2015evidence}. All data obtained from the SMS can be accessed using the following link: \href{https://bit.ly/4cTJ2gc}{https://bit.ly/4cTJ2gc}.

Once the data extraction form had been validated, we performed a full analysis of all the articles selected for this SMS. The extraction task was performed independently by  two researchers (the first author and a master's student from our research group). At the end, the collected data was consolidated into a spreadsheet. A third researcher, the second author, validated and assessed whether the extracted data effectively addressed the SMS research questions.

The process of analyzing and coding the data was performed using thematic analysis techniques guided by \cite{clarke2013successful} and \cite{diba2011recommended}. Thematic analysis is a systematic approach to identifying, analyzing, and reporting patterns (themes) in a set of qualitative data. In this approach, themes can be identified in a data-driven, \textit{bottom-up} way, on the basis of what is in the data; alternatively, they can be identified in a more \textit{top-down} fashion, where the researcher uses the data to explore particular theoretical ideas or brings those to bear on the analysis being conducted (bottom-up and top-down approaches are often combined in one analysis). In the process of identifying themes, we used both a bottom-up and top-down approach. In the \textit{top-down} approach, we exploit the concepts presented by \cite{mayer1995integrative}, Ability, Integrity, and Benevolence to initially classify the data. In the \textit{bottom-up} approach, we analyzed the data that could not be classified previously and created data-driven categories.

In the coding phase, we employed a complete systematic approach, encompassing two distinct types of codes: semantic (based on the semantic meaning of the data) and latent (codes derived from the researcher's conceptual and theoretical frameworks). By combining these codes, our analysis aimed to provide a holistic understanding of the dataset, capturing the extracted information’s explicit and implicit dimensions.

The coding process began immediately after the complete data extraction stages. Initially, we familiarized ourselves with all the extracted data by reading and re-reading the data to understand the content. 
We then performed the coding and assigned codes to each piece of extracted data. In this process, we coded the attributes that can influence interpersonal trust. 

The next step was to look for patterns in the codes, grouping the data into categories that reflected a broader idea. Finally, the categories were classified according to the literature on trust and given the following themes: Ability and Integrity. For the categories that could not be classified into these themes, we defined new themes to represent the categories according to the data extracted.


\section{Results}
This section presents an overview of studies and answers to the  RQs.

\subsection{\textbf{Articles Overview}}

This subsection provides an overview of the articles, guided by demographic aspects,  responses to research questions, and finally, based on the empirical strategies employed.

\begin{figure}[!ht]
\centering
\includegraphics[width=.8\textwidth]{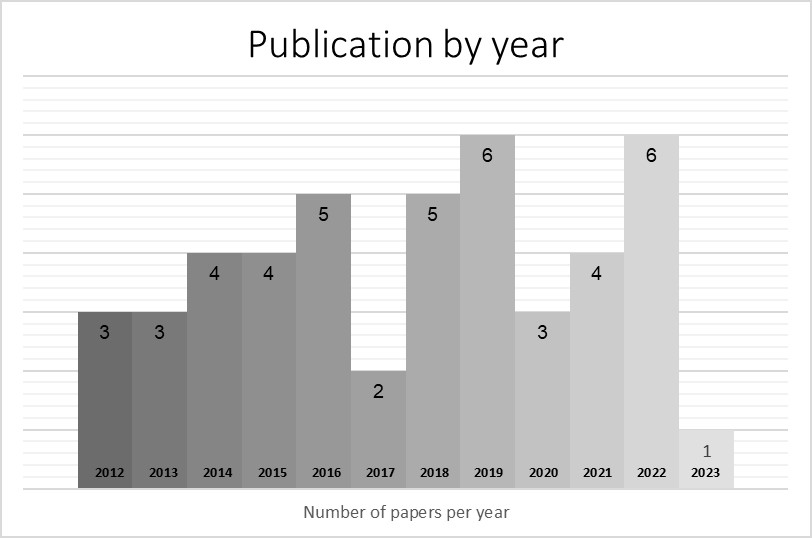}
\caption{Publication by year}
\label{fig:publication_year}
\end{figure}

Table \ref{tab:demographic_data} presents the 46 selected articles, indicating the year and the publication type (Pb, C=Conference, and J=Journal). An identification code is assigned to each article and used throughout this work. 26 articles were published in journals, and 20 articles were presented at conferences. The number of articles published per year remains relatively consistent throughout the period covered by the SMS (Figure \ref{fig:publication_year}). However, there is a slight variation in the period analyzed. The years 2019 and 2022 had the highest number of publications, with six each, and 2023 had the lowest number, with one publication

Considering RQ content-related issues, the articles were sorted according to the outcomes of each RQ. This resulted in the identification of five categories (see table \ref{tab:answers}): (i) Trust definitions: articles that provide explicit trust definitions along with analytical insights; (ii) Interpersonal attributes: articles in which we identified attributes that influence trust, fostering the development of trust among students; (iii) Acquiring trust: aspects that influence trust a positively reinforcing factor, leading one student to acquiring trust in another; (iv) Losing trust, articles shed light on aspects that harm trust, leading to a decline in trust among students; and (v) Supporting features, we identified implemented features that support trust aspects and processes among students.

\small
\footnotesize
\begin{longtable}{p{0.3cm} p{6.5cm} p{2.1cm} p{0.1cm} p{0.2cm}}
\caption{Demographic Data Table} \label{tab:demographic_data}
\\\hline
\textbf{\small Cod} & \multicolumn{1}{c}{\textbf{\small Title}} & \multicolumn{1}{c}{\textbf{\small Author}} & \multicolumn{1}{c}{\textbf{\small Pb}} & \multicolumn{1}{c}{\textbf{\small Year}} \\
\hline
\endfirsthead

\multicolumn{4}{c}{{\tablename} \thetable{} -- Continuation} \\
\hline
{\textbf{\small Cod}} & \multicolumn{1}{c}{\textbf{\small Title}}  & \multicolumn{1}{c}{\textbf{\small Author}} & \multicolumn{1}{c}{\textbf{\small Pb}} & \multicolumn{1}{c}{\textbf{\small Year}} \\
\hline
\endhead

\hline
\endfoot

\hline
\endlastfoot

\vspace{-0.1cm}\hypertarget{A1}{A1} &  \vspace{-0.1cm}Emotions and pair trust in asynchronous hospitality cultural exchange for students in Taiwan and Hong Kong & \vspace{-0.1cm}\cite{wang_2012_emotions_and_pair} & \vspace{-0.1cm}J & \vspace{-0.1cm} 2012\\


 \vspace{-0.1cm}\hypertarget{A2}{A2} & \vspace{-0.1cm}Trust evaluation model based on user trust cloud and user capability in E-learning service &  \vspace{-0.1cm}\cite{A2} & \vspace{-0.1cm}C &  \vspace{-0.1cm} 2012\\
 

\vspace{-0.1cm}\hypertarget{A3}{A3} & \vspace{-0.1cm}Trust, avatars, and electronic communications: Implications for e-learning & \vspace{-0.1cm} \cite{morrison_2012_trust_Avatar} & \vspace{-0.1cm}J &  \vspace{-0.1cm} 2012\\


\vspace{-0.1cm}\hypertarget{A4}{A4} & \vspace{-0.1cm}Effective trust-aware E-learning recommender system based on learning styles and knowledge levels & \vspace{-0.1cm}\cite{dwivedi_2013_effective_trust_ware} & \vspace{-0.1cm}J & \vspace{-0.1cm} 2013\\

\vspace{-0.1cm}\hypertarget{A5}{A5} & \vspace{-0.1cm}Toward dependable online peer assessments a concept of the trust manegement on peer assessments & \vspace{-0.1cm} \cite{fujihara_2013_toward_dependable_online} & \vspace{-0.1cm}C & \vspace{-0.1cm} 2013\\


\vspace{-0.1cm}\hypertarget{A6}{A6} & \vspace{-0.1cm}Towards a trust model in e-learning: Antecedents of a student's trust & \vspace{-0.1cm}\cite{wongse_2013_towards_a_trust_model} & \vspace{-0.1cm}C & \vspace{-0.1cm} 2013\\


\vspace{-0.1cm}\hypertarget{A7}{A7} & \vspace{-0.1cm}A collective intelligence approach for building student's trustworthiness profile in online learning & \vspace{-0.1cm}\cite{miguel2014collective} & \vspace{-0.1cm}C & \vspace{-0.1cm} 2014\\

\vspace{-0.1cm}\hypertarget{A8}{A8} & \vspace{-0.1cm}Building student trust in online learning environments & \vspace{-0.1cm}\cite{wang_2014_building_student_trust} & \vspace{-0.1cm}J & \vspace{-0.1cm} 2014\\

\vspace{-0.1cm}\hypertarget{A9}{A9} & \vspace{-0.1cm}Predicting Trustworthiness Behavior to Enhance Security in On-line Assessment & \vspace{-0.1cm}\cite{miguel_2014_predicting_trustworthiness} & \vspace{-0.1cm}C & \vspace{-0.1cm} 2014\\

\vspace{-0.1cm}\hypertarget{A10}{A10} & \vspace{-0.1cm}Security in Online Learning Assessment Towards an Effective Trustworthiness Approach to Support E-Learning Teams & \vspace{-0.1cm} \cite{miguel_2014_security_in_online} & \vspace{-0.1cm}C & \vspace{-0.1cm} 2014\\

\vspace{-0.1cm}\hypertarget{A11}{A11} & \vspace{-0.1cm}Social Analytics Framework to Boost Recommendation in Online Learning Communities & \vspace{-0.1cm}\cite{li_2015_social_analytics} & \vspace{-0.1cm}C & \vspace{-0.1cm} 2015\\

\vspace{-0.1cm}\hypertarget{A12}{A12} & \vspace{-0.1cm}Trust and forgiveness in virtual teams: A study in Algerian e-learning context & \vspace{-0.1cm} \cite{laifa_2015_trust_and_forgiveness} & \vspace{-0.1cm}C & \vspace{-0.1cm} 2015\\

\vspace{-0.1cm}\hypertarget{A13}{A13} & \vspace{-0.1cm}Trust evaluation for inter-organization knowledge sharing via the e-learning community & \vspace{-0.1cm}\cite{wu_2015_trust_evalueation} & \vspace{-0.1cm}J & \vspace{-0.1cm} 2015\\

\vspace{-0.1cm}\hypertarget{A14}{A14} & \vspace{-0.1cm}Trust-based community assessment & \vspace{-0.1cm}\cite{gutierrez_2015_trust_based} & \vspace{-0.1cm}J & \vspace{-0.1cm} 2015\\

\vspace{-0.1cm}\hypertarget{A15}{A15} & \vspace{-0.1cm}Bayesian Networks to predict reputation in Virtual Learning Communities & \vspace{-0.1cm}\cite{chamba_2016_bayesian_networks} & \vspace{-0.1cm}C & \vspace{-0.1cm} 2016\\

\vspace{-0.1cm}\hypertarget{A16}{A16} & \vspace{-0.1cm}How students regard trust in an elearning context & \vspace{-0.1cm}\cite{dwyer_2016_how_students_regard} & \vspace{-0.1cm}C & \vspace{-0.1cm} 2016\\

\vspace{-0.1cm}\hypertarget{A17}{A17} & \vspace{-0.1cm}The effect of online participation in online learning course for studying trust in information and communication technologies & \vspace{-0.1cm}\cite{widjaja_2016_the_effect_of} & \vspace{-0.1cm}J & \vspace{-0.1cm} 2016\\

\vspace{-0.1cm}\hypertarget{A18}{A18} & \vspace{-0.1cm}Trust-Aware Peer Assessment Using Multi-Armed Bandit Algorithms & \vspace{-0.1cm}\cite{chan_2016_trust_ware_peer} & \vspace{-0.1cm}C & \vspace{-0.1cm} 2016\\

\vspace{-0.1cm}\hypertarget{A19}{A19} & \vspace{-0.1cm}Web of Credit: Adaptive Personalized Trust Network Inference From Online Rating Data & \vspace{-0.1cm}\cite{mao_2016_web_of_credit} & \vspace{-0.1cm}J & \vspace{-0.1cm} 2016\\

\vspace{-0.1cm}\hypertarget{A20}{A20} & \vspace{-0.1cm}Online training for business plan writing through the World Café method: the roles of leadership and trust & \vspace{-0.1cm}\cite{chang_2017_online_training} & \vspace{-0.1cm}J & \vspace{-0.1cm} 2017\\

\vspace{-0.1cm}\hypertarget{A21}{A21} & \vspace{-0.1cm}Synergy Effect of the TeSLA Project in Management of Engineering Higher Education & \vspace{-0.1cm}\cite{tsankova_2017_synergy_effect} & \vspace{-0.1cm} C & \vspace{-0.1cm} 2017\\

\vspace{-0.1cm}\hypertarget{A22}{A22} & \vspace{-0.1cm}A trust-aware neural collaborative filtering for E-learning recommendation & \vspace{-0.1cm}\cite{deng_2018_trust_ware_neural} & \vspace{-0.1cm}J & \vspace{-0.1cm} 2018\\

\vspace{-0.1cm}\hypertarget{A23}{A23} & \vspace{-0.1cm}Effect of trust and perceived reciprocal benefit on students' knowledge sharing via facebook and academic performance & \vspace{-0.1cm}\cite{sedighehm_2018_effect} & \vspace{-0.1cm}J & \vspace{-0.1cm} 2018\\

\vspace{-0.1cm}\hypertarget{A24}{A24} & \vspace{-0.1cm}E-learning technology and higher education: the impact of organizational trust & \vspace{-0.1cm}\cite{boe_2018_learning_technology} & \vspace{-0.1cm}J & \vspace{-0.1cm} 2018\\

\vspace{-0.1cm}\hypertarget{A25}{A25} & \vspace{-0.1cm}Exploring antecedent factors toward knowledge sharing intention in E-learning & \vspace{-0.1cm}\cite{kunthi_2018_exploring_antecedent} & \vspace{-0.1cm}C & \vspace{-0.1cm} 2018\\

\vspace{-0.1cm}\hypertarget{A26}{A26} & \vspace{-0.1cm}In MOOCs we trust: Learner perceptions of MOOC quality via trust and credibility & \vspace{-0.1cm}\cite{costello_2018_In_moocs_we_trust} & \vspace{-0.1cm}J & \vspace{-0.1cm} 2018\\

\vspace{-0.1cm}\hypertarget{A27}{A27} & \vspace{-0.1cm}A close look at trust among team members in online learning communities & \vspace{-0.1cm}\cite{tseng_2019_a_close_look} & \vspace{-0.1cm}J & \vspace{-0.1cm} 2019\\

\vspace{-0.1cm}\hypertarget{A28}{A28} & \vspace{-0.1cm}A Comparative Analysis of OSN and SIoT Trust Models for a Trust Model Adapted to MOOCs Platforms & \vspace{-0.1cm}\cite{elghomary_2019_a_comparative_analysis} & \vspace{-0.1cm}C & \vspace{-0.1cm} 2019\\

\vspace{-0.1cm}\hypertarget{A29}{A29} & \vspace{-0.1cm}An efficient personalized trust based hybrid recommendation (TBHR) strategy for e-learning system in cloud computing & \vspace{-0.1cm}\cite{bhaskaran2019efficient} & \vspace{-0.1cm}J & \vspace{-0.1cm} 2019\\

\vspace{-0.1cm}\hypertarget{A30}{A30} & \vspace{-0.1cm}Dynamic Peer Recommendation System based on Trust Model for sustainable social tutoring in MOOCs & \vspace{-0.1cm}\cite{elghomary_2019_dynamic_peer} & \vspace{-0.1cm}C & \vspace{-0.1cm} 2019\\

\vspace{-0.1cm}\hypertarget{A31}{A31} & \vspace{-0.1cm}Integrating Student Trust in a Conceptual Model for Assessing Learning Management System Success in Higher Education: An Empirical Analysis & \vspace{-0.1cm}\cite{dorobuact_2019_integrating_student} & \vspace{-0.1cm}J & \vspace{-0.1cm} 2019\\

\vspace{-0.1cm}\hypertarget{A32}{A32} & \vspace{-0.1cm}The comparison of trust development in virtual and face-to-face collaborative learning groups & \vspace{-0.1cm}\cite{baturay_2019_the_comparison_of} & \vspace{-0.1cm}J & \vspace{-0.1cm} 2019\\

\vspace{-0.1cm}\hypertarget{A33}{A33} & \vspace{-0.1cm}Case Study of Collaborative Learning in a Massive Open Online Course & \vspace{-0.1cm}\cite{feng_2020_case_study} & \vspace{-0.1cm}C & \vspace{-0.1cm} 2020\\

\vspace{-0.1cm}\hypertarget{A34}{A34} & \vspace{-0.1cm}Improving the Quality of Learning Management System (LMS) based on Student Perspectives Using UTAUT2 and Trust Model & \vspace{-0.1cm}\cite{widjaja_2020_improving_the_quality} & \vspace{-0.1cm}C & \vspace{-0.1cm} 2020\\

\vspace{-0.1cm}\hypertarget{A35}{A35} & \vspace{-0.1cm}The role of e-learning service quality and e-trust on e-loyalty & \vspace{-0.1cm}\cite{pham_2020_the_role_of} & \vspace{-0.1cm}J & \vspace{-0.1cm} 2020\\

\vspace{-0.1cm}\hypertarget{A36}{A36} & \vspace{-0.1cm}Role of Design Attributes to Determine the Intention to Use Online Learning via Cognitive Beliefs & \vspace{-0.1cm}\cite{younas_2021_role_of_design} & \vspace{-0.1cm}J & \vspace{-0.1cm} 2021\\

\vspace{-0.1cm}\hypertarget{A37}{A37} & \vspace{-0.1cm}The Influence of e-WOM Information Characteristics on Learning Trust and e-WOM Intention Among Online Learning & \vspace{-0.1cm}\cite{noh_2021_the_influence_of} & \vspace{-0.1cm}J & \vspace{-0.1cm} 2021\\

\vspace{-0.1cm}\hypertarget{A38}{A38} & \vspace{-0.1cm}Trust in pandemic-induced online learning: Competitive advantage of closure and reputation & \vspace{-0.1cm}\cite{arnado_2021_trust_in_pandemic} & \vspace{-0.1cm}J & \vspace{-0.1cm} 2021\\

\vspace{-0.1cm}\hypertarget{A39}{A39} & \vspace{-0.1cm}Trusted cooperative e-learning service deployment model in multi-cloud environment & \vspace{-0.1cm}\cite{udhayakumar_2021_trusted_cooperative} & \vspace{-0.1cm}C & \vspace{-0.1cm} 2021\\

\vspace{-0.1cm}\hypertarget{A40}{A40} & \vspace{-0.1cm}A deep learning based trust- and tag-aware recommender system & \vspace{-0.1cm}\cite{ahmadian_2022_a_deep_learning} & \vspace{-0.1cm}J & \vspace{-0.1cm} 2022\\

\vspace{-0.1cm}\hypertarget{A41}{A41} & \vspace{-0.1cm}A Novel Precise Personalized Learning Recommendation Model Regularized with Trust and Influence & \vspace{-0.1cm}\cite{zhang_2022_a_novel_precise} & \vspace{-0.1cm}J & \vspace{-0.1cm} 2022\\

\vspace{-0.1cm}\hypertarget{A42}{A42} & \vspace{-0.1cm}Design of a Smart MOOC Trust Model: Towards a Dynamic Peer Recommendation to Foster Collaboration and Learner’s Engagement & \vspace{-0.1cm}\cite{Elghomary_Bouzidi_Daoudi_2022_design_of_smart_MOOC} & \vspace{-0.1cm}J & \vspace{-0.1cm} 2022\\

\vspace{-0.1cm}\hypertarget{A43}{A43} & \vspace{-0.1cm}Extended UTAUT Model to Analyze the Acceptance of Virtual Assistant’s Recommendations Using Interactive Visualizations & \vspace{-0.1cm} \cite{valtolina_2022_extended_utaut} & \vspace{-0.1cm}C & \vspace{-0.1cm} 2022\\

\vspace{-0.1cm}\hypertarget{A44}{A44} & \vspace{-0.1cm}Team satisfaction, identity, and trust: a comparison of face-to-face and virtual student teams & \vspace{-0.1cm}\cite{mayfield2022team} & \vspace{-0.1cm}J & \vspace{-0.1cm} 2022\\

\vspace{-0.1cm}\hypertarget{A45}{A45} & \vspace{-0.1cm}The effects of students perceived usefulness and trustworthiness of peer feedback on learning satisfaction in online learning environments & \vspace{-0.1cm}\cite{taghizadeh2022effects} & \vspace{-0.1cm}C & \vspace{-0.1cm} 2022\\

\vspace{-0.1cm}\hypertarget{A46}{A46} & \vspace{-0.1cm}T-VLC: A Trust Model for Virtual Learning Communities & \vspace{-0.1cm} \cite{chamba_tvlc2023} & \vspace{-0.1cm}J & \vspace{-0.1cm} 2023\\

\hline

\end{longtable}

\normalsize

Furthermore, as outlined in reference \cite{wohlin2012experimentation}, the empirical strategies utilized in the articles were categorized into three primary categories: survey, case study, and experiment. (see Table \ref{tab:article_proposal_empiric}). Among these approaches, experiments were identified as the most frequently employed empirical strategy in the selected articles.  Some articles did not apply any empirical strategy and thus were classified as no empirical evidence. The empirical strategies adopted in the articles can be seen in table \ref{tab:article_proposal_empiric}.

\small
\footnotesize
\begin{table}[!ht]
\centering

\begin{tabular}
{p{0.2\textwidth}>{\raggedright\arraybackslash}p{0.2\textwidth}>{\raggedright\arraybackslash}p{0.2\textwidth} p{0.24\textwidth}}

\hline
\textbf{Experiment} & \textbf{Survey} & \textbf{Case Study} & \textbf{No empirical evidence}\\ \hline
 \parbox[t]{0.4\textwidth}{\hyperlink{A16}{A16}, \hyperlink{A17}{A17}, \hyperlink{A27}{A27}, \\ \hyperlink{A32}{A32}, \hyperlink{A44}{A44}, \hyperlink{A45}{A45}, \\ \hyperlink{A9}{A9}, \hyperlink{A13}{A13}, \hyperlink{A19}{A19},\\ \hyperlink{A46}{A46},  \hyperlink{A4}{A4}, \hyperlink{A22}{A22}, \\ \hyperlink{A29}{A29}, \hyperlink{A30}{A30}, \hyperlink{A40}{A40}, \\ \hyperlink{A41}{A41}, \hyperlink{A14}{A14}, \hyperlink{A15}{A15}, \\ \hyperlink{A21}{A21}, \hyperlink{A36}{A36}, \hyperlink{A42}{A42}} & \parbox[t]{0.2\textwidth}{\hyperlink{A1}{A1}, \hyperlink{A23}{A23}, \hyperlink{A24}{A24}, \hyperlink{A25}{A25}, \hyperlink{A26}{A26}, \hyperlink{A34}{A34}, \hyperlink{A35}{A35}, \hyperlink{A37}{A37}, \hyperlink{A8}{A8}, \hyperlink{A12}{A12}, \hyperlink{A31}{A31}, \hyperlink{A43}{A43}} & \parbox[t]{0.2\textwidth}{\hyperlink{A3}{A3}, \hyperlink{A20}{A20}, \hyperlink{A33}{A33}, \hyperlink{A38}{A38}} & \hyperlink{A2}{A2}, \hyperlink{A5}{A5}, \hyperlink{A6}{A6}, \hyperlink{A7}{A7}, \hyperlink{A10}{A10}, \hyperlink{A11}{A11}, \hyperlink{A18}{A18}, \hyperlink{A28}{A28}, \hyperlink{A39}{A39}\\

21(46\%) & 12 (26\%) & 4 (9\%) & 9 (20\%) \\ \hline

\end{tabular}\\
\caption{Article proposals and empirical strategies}
\label{tab:article_proposal_empiric}

\end{table}

\normalsize

All articles accepted for analysis provided data relevant to at least one research question (Figure \ref{tab:answers}). In summary, 48\% (22) of the articles defined trust, 65\% (30) presented attributes that influence interpersonal trust, 63\% (29) presented attributes for acquiring trust, 13\% (6) presented attributes that can cause trust to be losing, and 39\% (18) presented trust-supporting features.

\small


\begin{table}[H]
\centering
\begin{tabular}{lp{9cm}}
\hline
\textbf{Answers obtained} & \textbf{Articles} \\
\hline
Trust definition & \hyperlink{A1}{A1}, \hyperlink{A3}{A3}, \hyperlink{A6}{A6}, \hyperlink{A7}{A7}, \hyperlink{A8}{A8}, \hyperlink{A12}{A12}, \hyperlink{A13}{A13}, \hyperlink{A15}{A15}, \hyperlink{A17}{A17}, \hyperlink{A20}{A20}, \hyperlink{A24}{A24}, \hyperlink{A27}{A27}, \hyperlink{A28}{A28}, \hyperlink{A30}{A30}, \hyperlink{A32}{A32}, \hyperlink{A34}{A34}, \hyperlink{A35}{A35}, \hyperlink{A36}{A36}, \hyperlink{A37}{A37}, \hyperlink{A39}{A39}, \hyperlink{A42}{A42}, \hyperlink{A44}{A44}, \hyperlink{A46}{A46} \\[0.5cm]

Interpersonal attributes & \hyperlink{A1}{A1}, \hyperlink{A2}{A2}, \hyperlink{A3}{A3}, \hyperlink{A5}{A5}, \hyperlink{A6}{A6}, \hyperlink{A7}{A7}, \hyperlink{A8}{A8}, \hyperlink{A9}{A9}, \hyperlink{A10}{A10}, \hyperlink{A11}{A11}, \hyperlink{A13}{A13}, \hyperlink{A14}{A14}, \hyperlink{A16}{A16}, \hyperlink{A18}{A18}, \hyperlink{A19}{A19}, \hyperlink{A20}{A20}, \hyperlink{A21}{A21}, \hyperlink{A23}{A23}, \hyperlink{A26}{A26}, \hyperlink{A27}{A27}, \hyperlink{A28}{A28}, \hyperlink{A30}{A30}, \hyperlink{A31}{A31}, \hyperlink{A32}{A32}, \hyperlink{A33}{A33}, \hyperlink{A34}{A34}, \hyperlink{A38}{A38}, \hyperlink{A40}{A40}, \hyperlink{A41}{A41}, \hyperlink{A42}{A42}, \hyperlink{A43}{A43}, \hyperlink{A44}{A44}, \hyperlink{A45}{A45}, \hyperlink{A46}{A46} \\[0.9cm]

Acquiring trust & \hyperlink{A1}{A1}, \hyperlink{A3}{A3}, \hyperlink{A5}{A5},\hyperlink{A7}{A7}, \hyperlink{A8}{A8}, \hyperlink{A10}{A10}, \hyperlink{A13}{A13}, \hyperlink{A14}{A14}, \hyperlink{A16}{A16}, \hyperlink{A18}{A18}, \hyperlink{A19}{A19}, \hyperlink{A20}{A20}, \hyperlink{A21}{A21}, \hyperlink{A22}{A22}, \hyperlink{A23}{A23}, \hyperlink{A26}{A26}, \hyperlink{A27}{A27}, \hyperlink{A28}{A28}, \hyperlink{A30}{A30}, \hyperlink{A32}{A32}, \hyperlink{A34}{A34}, \hyperlink{A38}{A38}, \hyperlink{A40}{A40}, \hyperlink{A41}{A41}, \hyperlink{A42}{A42}, \hyperlink{A44}{A44}, \hyperlink{A45}{A45}, \hyperlink{A46}{A46} \\[0.5cm]

losing trust & \hyperlink{A3}{A3}, \hyperlink{A9}{A9}, \hyperlink{A10}{A10}, \hyperlink{A28}{A28}, \hyperlink{A30}{A30}, \hyperlink{A46}{A46} \\[0.3cm]

Supporting trust & \hyperlink{A1}{A1}, \hyperlink{A3}{A3}, \hyperlink{A4}{A4}, \hyperlink{A7}{A7}, \hyperlink{A9}{A9}, \hyperlink{A10}{A10}, \hyperlink{A14}{A14}, \hyperlink{A15}{A15}, \hyperlink{A17}{A17}, \hyperlink{A23}{A23}, \hyperlink{A25}{A25}, \hyperlink{A27}{A27}, \hyperlink{A29}{A29}, \hyperlink{A30}{A30}, \hyperlink{A32}{A32}, \hyperlink{A36}{A36}, \hyperlink{A40}{A40}, \hyperlink{A41}{A41} \\
\hline
\end{tabular}
\caption{Articles per Trust Analyzed Aspects}
\label{tab:answers}
\end{table}

\normalsize

18 articles were classified as \textit{Trust analysis and assessment} (18), which identified, analyzed, or assessed the phenomenon of trust in VLEs as a predetermining factor in the validation of some kind of technology or use of technologies. These articles address a variety of objectives, such as: trust factors and knowledge sharing (\hyperlink{A23}{A23}, \hyperlink{A25}{A25}, \hyperlink{A38}{A38}), trust in virtual learning platforms (VLEs, MOOC, CMOOC, and LMS), evaluation of trust levels
 (\hyperlink{A17}{A17}, \hyperlink{A20}{A20}, \hyperlink{A24}{A24}, \hyperlink{A26}{A26}, \hyperlink{A32}{A32}, \hyperlink{A33}{A33}, \hyperlink{A34}{A34}, \hyperlink{A35}{A35}, \hyperlink{A37}{A37}), trust for collaboration in teams (\hyperlink{A27}{A27}, \hyperlink{A44}{A44}), and  trust and communication in online learning (\hyperlink{A1}{A1}, \hyperlink{A3}{A3}, \hyperlink{A16}{A16}, \hyperlink{A45}{A45}).

Article categorized as \textit{Models} (15) presents conceptual models aimed at supporting trust in VLEs. These models aim to analyze trust in various aspects such as: assessment (\hyperlink{A2}{A2}, \hyperlink{A10}{A10}, \hyperlink{A9}{A9}; \hyperlink{A13}{A13},  \hyperlink{A7}{A7}, \hyperlink{A18}{A18});  collaboration, (\hyperlink{A39}{A39}, \hyperlink{A8}{A8}, \hyperlink{A28}{A28}); conflict resolution, (\hyperlink{A12}{A12}); quality Assurance, (\hyperlink{A19}{A19}, \hyperlink{A6}{A6}, \hyperlink{A31}{A31}); management and estimation, (\hyperlink{A46}{A46}); And artificial intelligence usage (\hyperlink{A43}{A43}).

 The use of \textit{recommendation systems} was proposed by 7 articles. Trusted peer recommendation and peer-recommended content were identified (\hyperlink{A4}{A4}, \hyperlink{A11}{A11}, \hyperlink{A22}{A22}, \hyperlink{A30}{A30}, \hyperlink{A40}{A40}), make trusted recommendations for student collaboration (\hyperlink{A29}{A29}); recommendations for personalized learning content considering students with distinct learning styles (\hyperlink{A41}{A41}) provide accurate recommendations for personalized learning features and improve students' self-esteem.

5 articles were classified as \textit{Tools}, proposing some tool that could support trust in VLEs.  This category covers articles that, in addition to the conceptual model, propose algorithms, a prototype or the system's features: We identified assessment services based on trust, which suggests assessments to be corrected by students (Peer Review) in the context of MOOCs (\hyperlink{A14}{A14}); performed prediction of student reputation and implemented a plugin for this purpose on the MOODLE platform (\hyperlink{A15}{A15});  offering mutual trust among students, teachers and institutions through facial and voice recognition, analysis of writing style, typing rhythm and plagiarism analysis (\hyperlink{A21}{A21}); analysis of student trust based on the intention to use VLEs (\hyperlink{A36}{A36}); and, finally, designing a MOOC based on trustworthy and untrustworthy behavior (\hyperlink{A42}{A42}).

\begin{shadedbox}
 \textbf{Key Takeaways - Overview Articles}  
 \begin{itemize}
     \item  Studies on trust in VLEs have been conducted in several areas, aiming to improve interpersonal trust in several contexts in online courses, such as: (i) use of trustworthy peer recommendation systems,  (ii) analysis of the use of visual representations with the use of symbols and the impact on student trust, (iii)algorithms to suggest trustworthy student-assessors to perform peer assessment, (iv) use of biometrics for safe and trustworthy assessment in VLEs, as well as the (v) construction of several VLE models and prototypes focused on trust.
     
    \item Experiments and surveys  methods are preferred strategies for evaluation, considering factors related to interpersonal trust attributes and mechanisms for building trust among students.
    \item No selected work addressed the trust reconstruction. 
 \end{itemize}

\end{shadedbox}

\subsection{\textbf{RQ 1. What interpersonal trust characteristics are considered in virtual learning environment?}} 
In this section we present RQ 1, which aims to investigate the characteristics of interpersonal trust in VLEs. To achieve this goal, three auxiliary RQs were created to identify the trust definitions adopted in the articles (RQ1.1), the trust attributes in VLEs (RQ 1.2), and to identify which attributes of interpersonal trust influence changes in interpersonal trust (RQ 1.3).

\subsubsection{RQ 1.1 How is trust defined?}

Regarding RQ 1.1, we examined the definitions of trust and noted that 50\% (23 articles) explicitly provided definitions describing the trust concepts. These definitions revealed common trust relationships frequently discussed in the literature, primarily focusing on interpersonal and organizational contexts. In 12 articles (\hyperlink{A1}{A1}, \hyperlink{A3}{A3}, \hyperlink{A6}{A6}, \hyperlink{A7}{A7}, \hyperlink{A15}{A15}, \hyperlink{A20}{A20}, \hyperlink{A27}{A27}, \hyperlink{A28}{A28}, \hyperlink{A30}{A30}, \hyperlink{A32}{A32}, \hyperlink{A35}{A35}, \hyperlink{A44}{A44}), trust is addressed in interpersonal relationships. Although the focus of the articles was on interpersonal trust, some presented characteristics related to organizational trust in their definition. Of these, 04 articles (\hyperlink{A8}{A8}, \hyperlink{A34}{A34}, \hyperlink{A36}{A36}  \hyperlink{A37}{A37}) used a more appropriate definition for an organizational environment, and 07 articles (\hyperlink{A12}{A12}, \hyperlink{A13}{A13}, \hyperlink{A17}{A17}, \hyperlink{A24}{A24}, \hyperlink{A39}{A39}, \hyperlink{A42}{A42} and \hyperlink{A46}{A46}) used a definition that explores both types. The analysis also revealed various supplementary approaches that enhance the conceptualization of trust. Approximately 73\% of the article used definitions of trust based on existing literature. Notably, discordance or conflict is absent among these definitions; instead, they complement each other, resulting in a more comprehensive understanding of trust.

The table \ref{tab_definition} displays a selection of definitions taken from the chosen articles, exactly as they were presented in the respective articles. These definitions were chosen as examples of key words (underlined) that characterize trust in various domains. The other definitions extracted are organized in table \ref{tab:table_def} in the appendix of this text.

\small
\renewcommand{\arraystretch}{1.1}
\begin{longtable}{p{0.5cm} p{12cm}}

\hline
\textbf{Cod} & \textbf{Definition} \\
\hline
\endhead
 \vspace{-0.1cm}\hyperlink{A1}{A1} & \vspace{-0.1cm}A characteristic for collaboration where members believe in character, \uline{ability}, \uline{integrity}, familiarity and the morality of each other \\[0.5cm]

 \vspace{-0.1cm}\hyperlink{A3}{A3} & \vspace{-0.1cm}Trust, the willingness to depend on another. The perception that someone possesses \uline{attributes} that are \uline{beneficial} to the trustor \\[0.5cm]

\vspace{-0.1cm}\hyperlink{A7}{A7} & \vspace{-0.1cm}Is a particular level of the subjective probability with which an agent \uline{assesses} that another agent (or group of agents) will perform a particular \uline{action}, before the agent can \uline{monitor} such action (or independently of his capability ever to be able to monitor it) and in a \uline{context} in which it affects its own action \\[0.9cm]

 \vspace{-0.1cm} \hyperlink{A12}{A12} & \vspace{-0.1cm}Trust is defined as the \uline{belief} that the trustee will act in the \uline{best interests} of the trustor in a given \uline{situation}, even when controls are unavailable, and it may not be in the trustee’s \uline{best interests} to do so \\[0.5cm]

 \vspace{-0.1cm} \hyperlink{A15}{A15} & \vspace{-0.1cm}Trust is defined as the extent to which an individual has confidence and is willing to \uline{interact} with someone based on \uline{words}, \uline{actions}, and \uline{decisions} of others. \\[0.5cm]

 \vspace{-0.1cm}\hyperlink{A17}{A17}, \hyperlink{A24}{A24} & \vspace{-0.1cm}Trust as the willingness of a party to be vulnerable to the \uline{actions} of another party based on the \uline{expectation} that the other party will \uline{perform a particular action} important to the trustor, irrespective of the ability to \uline{monitor or control} that other party \\[0.8cm]

 \vspace{-0.1cm}\hyperlink{A35}{A35} & \vspace{-0.1cm}Trust is a psychological state, that is, willing to accept \uline{damage} based on \uline{expectations} of another person's \uline{intention or behavior} \\[0.5cm] 

 \vspace{-0.1cm}\hyperlink{A42}{A42} & \vspace{-0.1cm}A qualitative or quantitative property of a trustee, evaluated by a Trustor as a measurable \uline{belief}, subjectively or objectively, for a \uline{given task}, in a specific \uline{context}, for a specific period \\[0.5cm]

 \vspace{-0.1cm}\hyperlink{A44}{A44} & \vspace{-0.1cm}Psychological state willing to accept \uline{vulnerability} based on positive \uline{expectations} of or others \\[0.1cm]
\hline
\caption{Trust Definitions}
\label{tab_definition}

\end{longtable}

\normalsize

\begin{figure}[h]
    \centering
    \includegraphics[width=\linewidth]{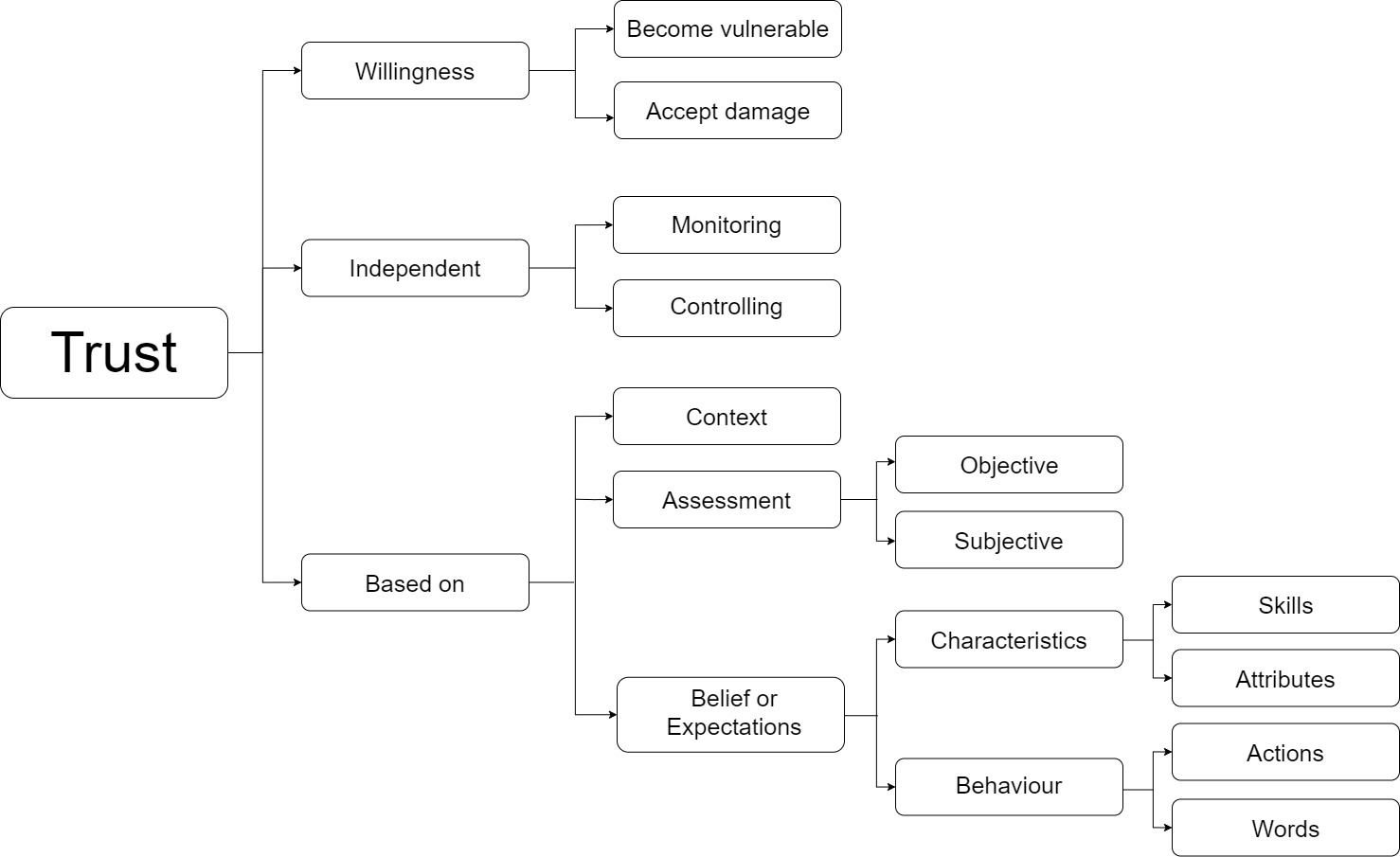}
    \caption{Conceptual Map of Trust Definition Elements }
    \label{fig:definition_trust}
\end{figure}

In the provided definitions, we identified terms frequently mentioned by the authors. Consequently, we extracted and organized these terms to enhance the understanding of the subject. Table \ref{tab_definition} highlights the words used by various authors that align with the domain definition studied in this SMS. Additionally, Figure \ref{fig:definition_trust} illustrates the characteristics of trust identified in the definitions.

Regarding the trust definitions, a common element is
\textit{Willingness} to trust (\hyperlink{A3}{A3}, \hyperlink{A6}{A6}, \hyperlink{A8}{A8}, \hyperlink{A9}{A9}, \hyperlink{A15}{A15}, \hyperlink{A17}{A17}, \hyperlink{A24}{A24}, \hyperlink{A32}{A32}, \hyperlink{A34}{A34}, \hyperlink{A35}{A35}, \hyperlink{A44}{A44})  that refers to the trustor innate level and propensity to trust others. This willingness also extends to making oneself vulnerable, taking risks, and accepting potential damages, which involves embracing the inherent risks in trust relationships, potentially putting something significant at risk, and being prepared to accept negative consequences if trust terms are not met.

Moreover, trust involves \textit{beliefs} (\hyperlink{A8}{A8}, \hyperlink{A30}{A30}, \hyperlink{A37}{A37}, \hyperlink{A35}{A35}, \hyperlink{A27}{A27}, \hyperlink{A28}{A28}, \hyperlink{A12}{A12}, \hyperlink{A13}{A13}, \hyperlink{A36}{A36}, \hyperlink{A42}{A42}), the positive attitude of the trustor, with the conviction that the trustee will honor the trust placed. Trust also involves  \textit{expectations} (\hyperlink{A35}{A35}, \hyperlink{A39}{A39}, \hyperlink{A24}{A24}, \hyperlink{A17}{A17}, \hyperlink{A44}{A44}, \hyperlink{A46}{A46}), which involves the hope that the trustee will fulfill that trust. In this sense, one party trusts the other to act in a beneficial, honest, competent, and predictable manner. 

Trust is considered to be \textit{contextual} (\hyperlink{A9}{A9}, \hyperlink{A42}{A42}, \hyperlink{A12}{A12}, \hyperlink{A16}{A16}), which depends on a \textit{qualitative or quantitative, objective or subjective} assessment by the trustor (\hyperlink{A42}{A42}), to place oneself in a situation of \textit{vulnerability} (\hyperlink{A24}{A24}, \hyperlink{A17}{A17}, \hyperlink{A44}{A44}) or to accept \textit{damage} (\hyperlink{A35}{A35}), based on \textit{beliefs} (\hyperlink{A8}{A8}, \hyperlink{A30}{A30}, \hyperlink{A37}{A37}, \hyperlink{A35}{A35}, \hyperlink{A27}{A27}, \hyperlink{A28}{A28}, \hyperlink{A12}{A12}, \hyperlink{A13}{A13}, \hyperlink{A36}{A36}, \hyperlink{A42}{A42}) or \textit{expectations} (\hyperlink{A35}{A35}, \hyperlink{A39}{A39}, \hyperlink{A24}{A24}, \hyperlink{A17}{A17}, \hyperlink{A44}{A44}, \hyperlink{A46}{A46} on \textit{actions} (\hyperlink{A30}{A30}, \hyperlink{A34}{A34}, \hyperlink{A24}{A24}, \hyperlink{A15}{A15}, \hyperlink{A17}{A17}), \textit{intentions} (\hyperlink{A35}{A35}), \textit{words} (\hyperlink{A15}{A15}), \textit{skills} (\hyperlink{A1}{A1}, \hyperlink{A23}{A23}), and \textit{attributes} (\hyperlink{A3}{A3}), \textit{independent of monitoring and control} structures (\hyperlink{A38}{A38}, \hyperlink{A24}{A24}, \hyperlink{A12}{A12}, \hyperlink{A17}{A17}, \hyperlink{A7}{A7}), with the aim of obtaining \textit{good results} (\hyperlink{A28}{A28}, \hyperlink{A6}{A6}, \hyperlink{A30}{A30}).

 The decision to trust something or someone starts with assessing the risks (\hyperlink{A17}{A17}, \hyperlink{A24}{A24}, \hyperlink{A35}{A35}, \hyperlink{A44}{A44}) and rewards (\hyperlink{A6}{A6}, \hyperlink{A8}{A8}, \hyperlink{A28}{A28}, \hyperlink{A30}{A30}, \hyperlink{A44}{A44}) involved. The trust relationship is based on the trustor needs (\hyperlink{A6}{A6}, \hyperlink{A8}{A8}), which result in mutual benefits when met by the trustee. On the other hand, failure to meet the trustor needs can cause damage and reduce trust levels. Therefore, the trustor analyzes whether the potential benefits of the relationship outweigh the associated risks.

Regarding assessment aspects, whether something or someone is worthy of trust, it is essential to consider the \textit{context} when evaluating a person's abilities (\hyperlink{A7}{A7}, \hyperlink{A9}{A9}, \hyperlink{A42}{A42}). Likewise, a person may be trustworthy for one task rather than for another. Moreover, assessment may be objective or subjective.  Subjective assessment does not require structures to generate trust (\hyperlink{A7}{A7}, \hyperlink{A42}{A42}), and it can occur through the trustor perception of some characteristic that the trustee possesses that inspires trust (\hyperlink{A1}{A1}, \hyperlink{A3}{A3}, \hyperlink{A20}{A20}, \hyperlink{A35}{A35}). E.g., by observing someone driving a car, the trustor can trust the driver to take them from one place to another. By perceiving that the driver operates the vehicle with skill and responsibility, the trustor feels confident about this trustee's skills. 
On the other hand, objective assessment (\hyperlink{A42}{A42}) results from the consideration of structures that ensure trust. E.g., a certain company needs car drivers and asks those interested for a driving license. It is assumed that a driver with a license issued by a competent body has the necessary skills to operate a vehicle. In this case, the assessment is based on formal criteria and official documentation.

Regarding the way the trustor assesses the trustee as trustworthy, (\hyperlink{A7}{A7}, \hyperlink{A20}{A20}, \hyperlink{A42}{A42}), the trustor performs an assessment based on a subjective or objective analysis of whether the trustee will take the appropriate measures to obtain good results. The trustor performs quantitative or qualitative assessments to determine whether the trustee can develop appropriate strategies to obtain the best result.

With regard to the trustor \textit{belief} (\hyperlink{A8}{A8}, \hyperlink{A12}{A12}, \hyperlink{A13}{A13}, \hyperlink{A27}{A27}, \hyperlink{A28}{A28}, \hyperlink{A30}{A30}, \hyperlink{A35}{A35}, \hyperlink{A36}{A36}, \hyperlink{A37}{A37}, \hyperlink{A42}{A42}), in the context of trust, these \textit{beliefs} are grounded in positive attitude of the trustor, reflecting their trust in the trustee's genuine motivations. E.g., student A trusts student B to perform a task because they believe in student B's capabilities and abilities. Trust is also influenced by the skills, competencies, and characteristics of the trustee.

Regarding \textit{expectations} \hyperlink{A35}{A35}, \hyperlink{A39}{A39}, \hyperlink{A24}{A24}, \hyperlink{A17}{A17}, \hyperlink{A44}{A44}, \hyperlink{A46}{A46}), on the other hand, they relate to the trustor willingness and hope that the trustee will act as expected. E.g., student A trusts student B with the \textit{expectation} that student B will be able to help student A achieve good results. Both \textit{belief} and \textit{expectations} are deeply influenced by the actions that the trustor foresees or expects the fiduciary agent to perform.

As far as the \textit{context} (\hyperlink{A7}{A7}, \hyperlink{A9}{A9}, \hyperlink{A42}{A42}) is concerned, trust is shaped by situational circumstances, the trustor specific needs, and the temporal aspect of trust formation. Trust depends on the \textit{context} in which the trustor and trustee are involved.

With regard to the need for supervision, the literature points out, based on the definitions analyzed, that trust does not need formal structures of \textit{controlling} (\hyperlink{A7}{A7}, \hyperlink{A12}{A12}, \hyperlink{A17}{A17}, \hyperlink{A24}{A24}) to regulate behavior, actions or performance and \textit{monitoring} (\hyperlink{A7}{A7}, \hyperlink{A17}{A17}, \hyperlink{A24}{A24}) to supervise actions. \textit{Controlling} involves adapting to the desired behavioral pattern, while\textit{ monitoring} involves activities such as observing and recording work behavior, with or without technology. Trustor can use standardized rules and procedures, as well as technological tools, to detect inappropriate conduct and record working hours.

The decision to place trust in another student is influenced by the educational context and the need of the trustor. This need often drives collaboration among students since the student trustor seeks to achieve fruitful results through this collaboration.

As a summary, trust can be seen as \textit {"Interpersonal Trust delineates the trustor innate inclination and predisposition towards reliance on others, based on the \uline{belief} or \uline{expectation} and on a \uline{subjective} or \uline{objective} assessment that the trustee will act according to what is expected in \uline{actions}, \uline{abilities} or \uline{behavior} in a given \uline{context}, \uline{independent of control} or \uline{monitoring structures} to achieve \uline{good outcomes}"}.

 \begin{shadedbox}
     \textbf{RQ1.1 Key Takeaways:} 
\begin{itemize}
    \item There are no significant differences in the definitions of trust used in the context of VLEs.
 \item The various approaches to VLEs definitions explore a variety of aspects, broadening the understanding and scope of the concept of trust in this area.
   \item The analysis of definitions reveals that trust is based on beliefs or expectations. Belief implies a positive attitude from the trustor, convinced that the trustee will honor the entrusted trust. On the other hand, expectations involve the hope that the trustee will fulfill the trust. 

    \item The foundation of trust, whether rooted in beliefs or expectations, lies in the trusted object of the trustee, which may encompass \textit{actions, intentions, behaviors, competencies, skills, or attributes}. In this context, the trustee can be an individual, a group, an organization, or even a technology.
     \item In online courses, the trust of the student trustor is based on cognitive trust and is intrinsically linked to the expectations of the student trustee. 
    
    \item Expectations are based on the transparency of the attitudes, i.e., acting as expected,  integrity, and competencies that the student trustor perceives in the student trustee, assessing their capability (or attributes) to be useful in a specific context.
\end{itemize}

 \end{shadedbox}



\subsection{RQ 1.2 What types of interpersonal trust attributes are considered in Virtual Learning Environments?}

In this section, we present and discuss personal attributes that support and affect trust in VLEs. As stated before in the methodology section \ref{sec:research}, we used the thematic analysis techniques to summarized data from the articles. 

Initially, data related to the characteristics of the trustee and trustor or associated with the trust phases were explicitly captured from articles texts and entered into a spreadsheet. 
We then used the concepts presented by \cite{mayer1995integrative} as predefined themes: ABILITY, BENEVOLENCE, and INTEGRITY as guidance to group extracted data.  We did not find data related to BENEVOLENCE as a factor in inducing trust in virtual environments. This may reflect that the judgment about the extent to which the trustee would act in the trustor best interests (i.e., BENEVOLENCE) may be difficult to assess in the virtual context, as this assessment requires a more intimate interpersonal relationship \citep{choi2019mechanism}. To see the attributes, classification into categories, and themes.

While  ABILITY and INTEGRITY are directly related to the trustor perception of the trustee aspects (e.g., behavior, skills, moral and ethical values), other interpersonal trust factors can be influenced by some aspects that go beyond the dyadic relationship of these two agents, being situated in a broader social space that connects individuals \citep{mcevily2021network}. Therefore, two new themes emerged: AFFINITY and NON-PERSONAL FACTORS as these attributes that are not directly related to just only trustees. In this case, AFFINITY groups together attributes of similarity of ideas, behaviors and the sharing of common friends in a trusted network. Regarding NON-PERSONAL FACTORS refer to external or contextual influences that affect the relationship without being directly related to the individual characteristics of the people involved.

Analyzing the data extracted, we identified 37 attributes of interpersonal trust presented in 32 articles and organized them into four themes. Each attribute is associated with a category and each category is associated with a theme. For better understanding, we have formatted the words to improve clarity and provide better understanding: themes have been formatted in uppercase, categories in bold and attributes in italics. 

\begin{figure}[H]
\includegraphics[width=\textwidth]{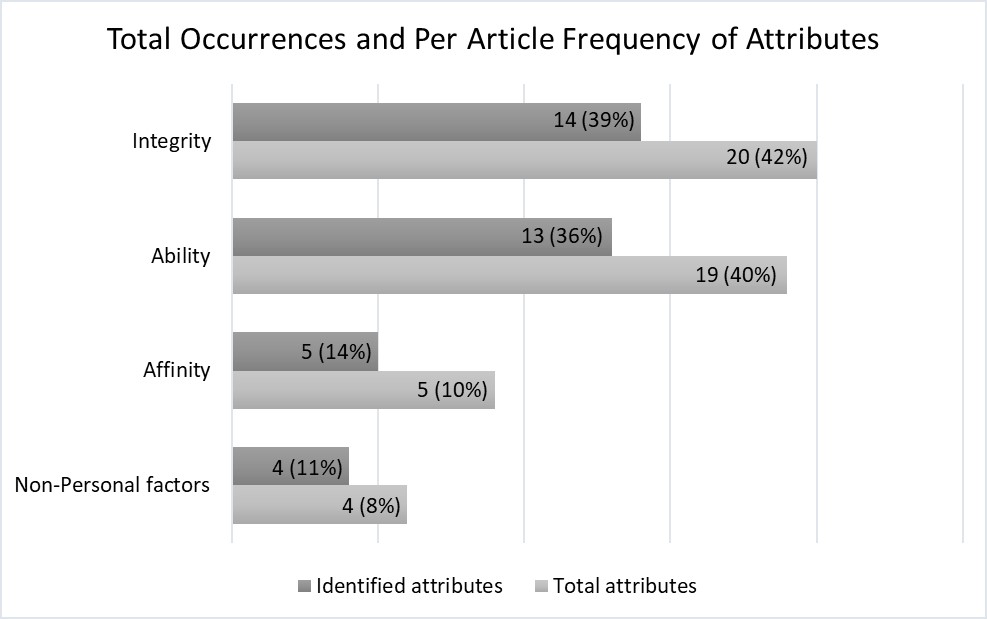}
\caption{Motivating factors for interpersonal trust in VLEs}
\label{fig:categories_interpersonal_trust}
\end{figure}

In this SMS we have identified ABILITY, INTEGRITY, AFFINITY and NON-PERSONAL FACTORS as themes. INTEGRITY and ABILITY are the themes that had the highest number of attributes identified in the articles. Figure \ref{fig:categories_interpersonal_trust}  shows the number of attributes identified per theme (dark gray) and the occurrences in the articles (light gray). E.g., INTEGRITY groups 14 (39\%) attributes identified in 20 (42\%) articles having the highest number of attributes occurring (See \hyperlink{DATA}{Dataset}).

\begin{figure}[H]
\includegraphics[width=.9 \textwidth]{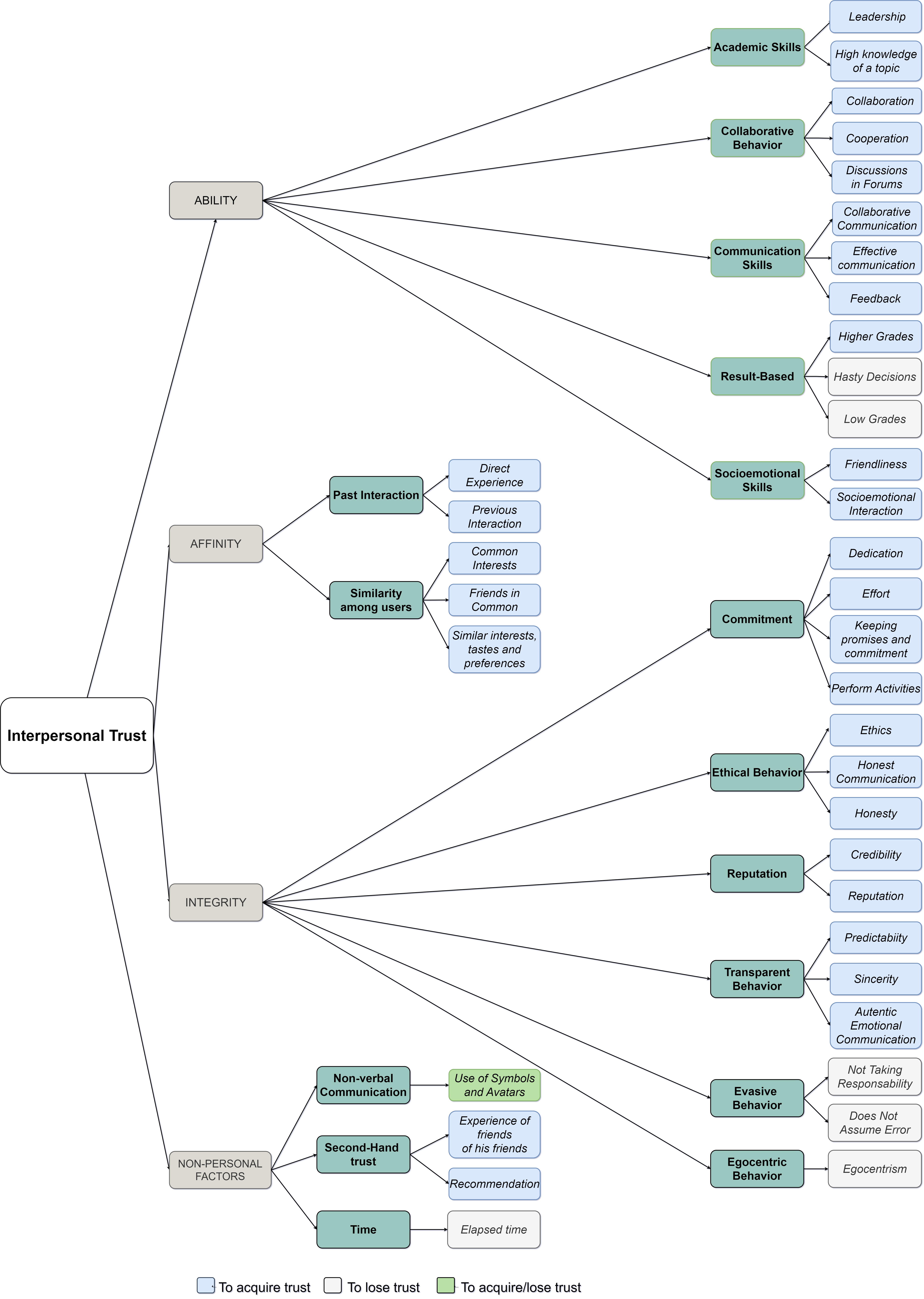}
\caption{Conceptual map of the attributes that influence interpersonal trust in VLEs}
\label{fig:Map}
\end{figure}

Moreover, in order to provide an overview of the attributes that influence trust, we elaborated a conceptual map that illustrates all the attributes identified. Figure \ref{fig:Map} presents the conceptual map in which attributes painted in blue are related to acquiring trust phase, light gray are related to losing trust (trust phases will be discussed in Section \ref{sec:rq1.3}, related to RQ 1.3). In the following paragraphs, we present the themes, categories, and attributes.

ABILITY refers to skills and competencies enabling an individual or entity to influence a specific domain \citep{mayer1995integrative}. E.g., trustee may possess high competence in a technical area, earning them trust in tasks related to that field. In VLEs, ABILITY groups is divided into five categories: \textbf{academic skills}  (relating to students' knowledge or skills that have been or can be acquired through in formal education instruction; \hyperlink{A20}{A20},  \hyperlink{A46}{A46}),  \textbf{collaborative behavior} (behaviors that lead to collaboration among students; \hyperlink{A16}{A16}, \hyperlink{A28}{A28}, \hyperlink{A10}{A10}, \hyperlink{A13}{A13}, \hyperlink{A30}{A30}), \textbf{communication skills}  (capability to transmit and understand information clearly and effectively; \hyperlink{A14}{A14}, \hyperlink{A27}{A27}, \hyperlink{A30}{A30}, \hyperlink{A32}{A32},   \hyperlink{A45}{A45}, \hyperlink{A46}{A46}), \textbf{result-based}  (attributes that influence trustor perception based on trustee academic results; \hyperlink{A10}{A10}, \hyperlink{A46}{A46}), and \textbf{socio-emotional skills}  (capability to relate to oneself and to others, showing social initiative and control of one's emotions in relationships with others, modulating behavioral and emotional reactivity in social interactions; \hyperlink{A1}{A1}, \hyperlink{A23}{A23}).

Attributes related to \textbf{academic skills}  include \textit{high knowledge on a given topic} (which refers to students' proficiency in a specific subject; \hyperlink{A46}{A46}), and \textit{leadership} (regarding the capability to lead and influence others in a positive way; \hyperlink{A20}{A20}).
\textbf{Collaborative behavior}  groups the following attributes \textit{collaboration} and \textit{cooperation} (relate to working together to achieve common goals; \hyperlink{A16}{A16}, \hyperlink{A28}{A28}, \hyperlink{A27}{A27}, \hyperlink{A30}{A30}, \hyperlink{A42}{A42},
\hyperlink{A6}{A6}, \hyperlink{A10}{A10}, \hyperlink{A13}{A13}, \hyperlink{A30}{A30}) and  \textit{discussions in forums}  (actively participating in ideas debates and exchanges; \hyperlink{A7}{A7}).
Regarding \textbf{{communication skills}}, we identified the communication attributes that can influence trust. The following three attributes were identified: \textit{collaborative communication}  (capability to communicate with other students in an inclusive way in order to achieve collective goals; \hyperlink{A27}{A27}); \textit{effective communication} (capability to communicate effectively and without noise; \hyperlink{A32}{A32}); and \textit{feedback in interactions} (providing constructive feedback in students interactions; \hyperlink{A14}{A14}, \hyperlink{A30}{A30}, \hyperlink{A45}{A45}, \hyperlink{A46}{A46}).

The \textbf{result-based} category is related to results produced by the trustee. In this category, we identified the following three attributes: \textit{higher grades} (indicative of good results in performed activities; \hyperlink{A18}{A18}, \hyperlink{A46}{A46}), \textit{hasty decisions} (based on negative results generated from hasty or rushed decisions; \hyperlink{A10}{A10}), and \textit{low grades} (indicative of unsatisfactory results; \hyperlink{A46}{A46}).
Finally, regarding \textbf{socio-emotional skills} category, we identified two attributes: \textit{friendliness} (to establish positive social interactions; \hyperlink{AA1}{A1}); and \textit{socio-emotional interaction}  (is related to emotional interactions between individuals, involving the ability to understand, interpret and respond to the emotions of others, as well as expressing one's own emotions appropriately; \hyperlink{A23}{A23}).

With regard to AFFINITY, it is a motivating factor for trust among students that shares several commonalities. This theme groups together the following categories: 
 \textbf{similarity among users} is a characteristic of identification among students, is governed by the following attributes: \textit{similar interests, tastes, and preferences} (when students have similar characteristics and tastes; \hyperlink{A30}{A30}, \hyperlink{A40}{A40}) and \textit{friends in common} (which suggests the possibility of trust arising from a mutual friend; \hyperlink{A22}{A22}, \hyperlink{A42}{A42}), and \textbf{past interactions} (trust that can arise from moments of interaction during academic activities or informal moments among students; \hyperlink{A14}{A14}, \hyperlink{A30}{A30}). In this category, we identified the following attributes: \textit{experience direct} (direct contact among students to perform collaborative activities; \hyperlink{A46}{A46}) and \textit{previous interact} (contact among students resulting from several direct collaboration activities, such as group work, or indirect through the use of discussion forums, or informal contacts arising in the VLE itself or outside it; \hyperlink{A1}{A1}, \hyperlink{A14}{A14}, \hyperlink{A30}{A30}).

 Regarding the INTEGRITY theme, it is related to the trustee-trustor relationship, in which the trustor perception is that the trustee adheres to a set of principles that are considered acceptable. In addition to this, the consistency of past actions, trustworthy communications, the trustor belief that the trustee possesses a strong sense of justice, and the extent to which the trustee actions are congruent with their words collectively influence the degree to which the trustee is judged to have integrity \citep{mayer1995integrative}. In VLEs, a student's perception of INTEGRITY can be influenced by \textbf{commitment} (relates to the commitment shown by students in relation to the goals, objectives, or responsibilities they have in the course; \hyperlink{A10}{A10}, \hyperlink{A21}{A21}, \hyperlink{A32}{A32}), 
\textbf{egocentric behavior} (self-focused behavior, based on the promotion of one's own interests; \hyperlink{A10}{A10}), 
\textbf{ethical behavior} (involve moral conduct, ethical principles and socially accepted norms of behavior; \hyperlink{A10}{A10}, \hyperlink{A23}{A23}, \hyperlink{A30}{A30}, \hyperlink{A34}{A34},), 
\textbf{evasive behavior} (tendency to avoid or evade responsibility; \hyperlink{A10}{A10}), 
\textbf{reputation}  (general perception that others have of the trustee based on their actions, behavior, and past performance; \hyperlink{A5}{A5}, \hyperlink{A6}{A6}, \hyperlink{A8}{A8}, \hyperlink{A11}{A11}, \hyperlink{A26}{A26}, \hyperlink{A28}{A28}, \hyperlink{A38}{A38}, \hyperlink{A46}{A46}), and 
\textbf{transparent behavior} (indicate clarity and openness in the trustee communication and actions, acting in a predictable manner; \hyperlink{A41}{A41}). Regarding  \textbf{commitment}, we identified four attributes: \textit{dedication} (to work hard towards a particular goal or task, showing persistence; \hyperlink{A32}{A32});
\textit{effort} (investing energy and time to achieve goals, showing determination and a willingness to overcome challenges; \hyperlink{A32}{A32}); \textit{keeping promises and agreements}  (keeping promises and honoring commitments, showing trustworthiness, responsibility and integrity; \hyperlink{A10}{A10}); and \textit{Perform activities}  (Perform out tasks and evaluative activities, with the aim of performing and completing activities properly; \hyperlink{A21}{A21}). Moreover, we classified attributes involving moral conduct, ethical principles, and socially accepted norms of behavior under \textbf{ethical behavior} category. In this category, we identified three attributes: \textit{ethics} (moral values that imply acting in accordance with certain principles; \hyperlink{A23}{A23}); and \textit{honesty} (being truthful and sincere in all interactions and communications; \hyperlink{A30}{A30}, \hyperlink{A34}{A34}); \textit{honest communication} (communication without information distortion; \hyperlink{A10}{A10}).

Some attributes were classified as \textbf{transparent behavior} category to indicate clarity and openness in the trustee communication and actions, acting in a predictable manner. The attributes relating to this category are: \textit{authentic emotional communication}   (willingness to express emotions genuinely and sincerely, allowing for a deeper and more meaningful connection with others; \hyperlink{A44}{A44});
\textit{predictability} (indicates that a person's actions and reactions are relatively stable; \hyperlink{A41}{A41}); and
\textit{sincerity} (transparency and frankness in communications; \hyperlink{A32}{A32}).
Regarding \textbf{reputation}, we classify attributes according to the general perception that others have of the student based on their actions, behavior, and past performance. In this category, we identified the student's \textit{credibility} (capability to be true and authentic; \hyperlink{A11}{A11}, \hyperlink{A26}{A26}) in the eyes of their peers.

With regard to self-focused behavior, based on egocentric attitudes and the promotion of one's own interests, we classify it as \textbf{egocentric behavior} category. We identified \textit{egocentrism} (self-focused attitudes and behaviors; \hyperlink{A10}{A10}) as an attribute. 
As for \textbf{evasive behavior} category, i.e., the tendency to avoid or evade responsibility, we identified the following attributes: \textit{does not assume error};  (reluctance or refusal to admit or acknowledge mistakes made; \hyperlink{A10}{A10}); and \textit{not taking responsibility} (behavior of denying responsibility; \hyperlink{A10}{A10}).

As stated before, some factors that influence trust discussed in the articles are independent of a dyadic relationship between trustor and trustee. In this perspective, we have identified some categories that do not need a direct relationship to influence trust among students: In the \textbf{time} category, we observed that \textit{elapsed time} (\textbf{elapsed time} between interactions;  \hyperlink{A9}{A9}, \hyperlink{A28}{A28}, \hyperlink{A30}{A30}) influences trust, particularly as interpersonal interaction decreases. In such instances, the shorter the interaction or collaboration \textit{elapsed time}, the greater the decline in trust. 
\textbf{Non-verbal communication} category, we identified the \textit{use of symbols and avatars} as being able to influence trust (\textit{use of symbols and avatars} in profiles can influence the trust of other students according to the context and how they understand the symbolism; \hyperlink{A3}{A3}). \textbf{Second-hand trust} category is based on what we learn about a person indirectly through our interactions with other people who have also had experiences with the same person and on the surrounding conditions that affect the social dynamics of trust \citep{mcevily2021network}. In this category, we identified \textit{experience of friends of his friends} (allows a student to use the successful experience of another student to relate to a third student, i.e., positive experience of a trustor friend with another person can influence the development of the trustor trust in that person; \hyperlink{A30}{A30}), and \textit{recommendation} (in VLEs occurs with an indication of someone or technology based on similar characteristics among users or according to the trustor interest in a specific skill; \hyperlink{A28}{A28}, \hyperlink{A30}{A30}) as attributes.

As an overview, 37 attributes were identified that can influence trust among students, arranged in four thematic areas. INTEGRITY, with 42\%, was the theme in which we identified the most interpersonal trust attributes, while ABILITY had 40\% of the attributes. Regarding AFFINITY among students, we identified 10\% of the attributes and NON-PERSONAL FACTORS 8\%.

As a highlight, the articles reveal a limited focus on evaluating trust attributes during a course. An exception is \hyperlink{A3}{A3}, which examines the influence \textit{use of symbols and avatars} on user trust, noting that the perception of trust can vary depending on the context. However, this study does not directly address the impact of actions and behaviors on the relationship between students or how these actions influence trust.

In summary, interpersonal trust is mainly related to INTEGRITY and students' ABILITY.  Regarding INTEGRITY, the attributes are mainly related to \textbf{commitment} and \textbf{ethical behavior}, while the ABILITY attributes are mainly related to \textbf{communication skills}, \textbf{collaborative behavior} and the results (\textbf{result-based}) obtained by the student. As for \textbf{similarity among users}, this is a relevant feature of AFFINITY that influences trust in VLEs. Furthermore, interpersonal trust among students can be influenced by NON-PERSONAL FACTORS, i.e., attributes that are not directly linked to the relationship between two students.

\begin{shadedbox}
\textbf{RQ1.2 Key Takeaways:} 
\begin{itemize}

\item Most of the articles (30) address some attribute related to trust.
\item Articles focus on the identification, conceptualization and implementation of attributes, but do little to assess students' perceptions and experiences.
\item ABILITY and INTEGRITY, themes proposed by  \cite{mayer1995integrative}, is the most cited attributes among the articles.
\item ABILITY is a fundamental personal characteristic for carrying out academic activities
\item INTEGRITY is necessary for initiating relationship of trust.
\item Two news themes AFFINITY and NON-PERSONAL FACTORS arose in order to group attributes  that go beyond the relationship of these two agents, being situated in a broader social space that connects individuals.

\item Attributes associated with negative aspects that lead to trust decrease have little attention.

\end{itemize}

\end{shadedbox}




\subsubsection{\textbf{RQ 1.3 In which evolutionary phases of trust can students interpersonal trust attributes be classified?}} \label{sec:rq1.3}

Considering the trust phases proposed by \cite{currall2003fragility} and \cite{fachrunnisa2010state} (acquiring, maintain, losing, and restore), we conducted our data extraction and analysis of selected articles. Thus, we observed the synonymous terms used, such as 'acquiring trust', 'acquisition trust', 'building trust', and 'establishing trust'  to identify the phases.   To classify the attributes in each phase, we examined the articles for influence indications of each attribute on a specific trust phase. In addition, at least one article had to associate the attribute with a phase.

\begin{figure}[h]
    \centering
    \includegraphics[width=1\linewidth]{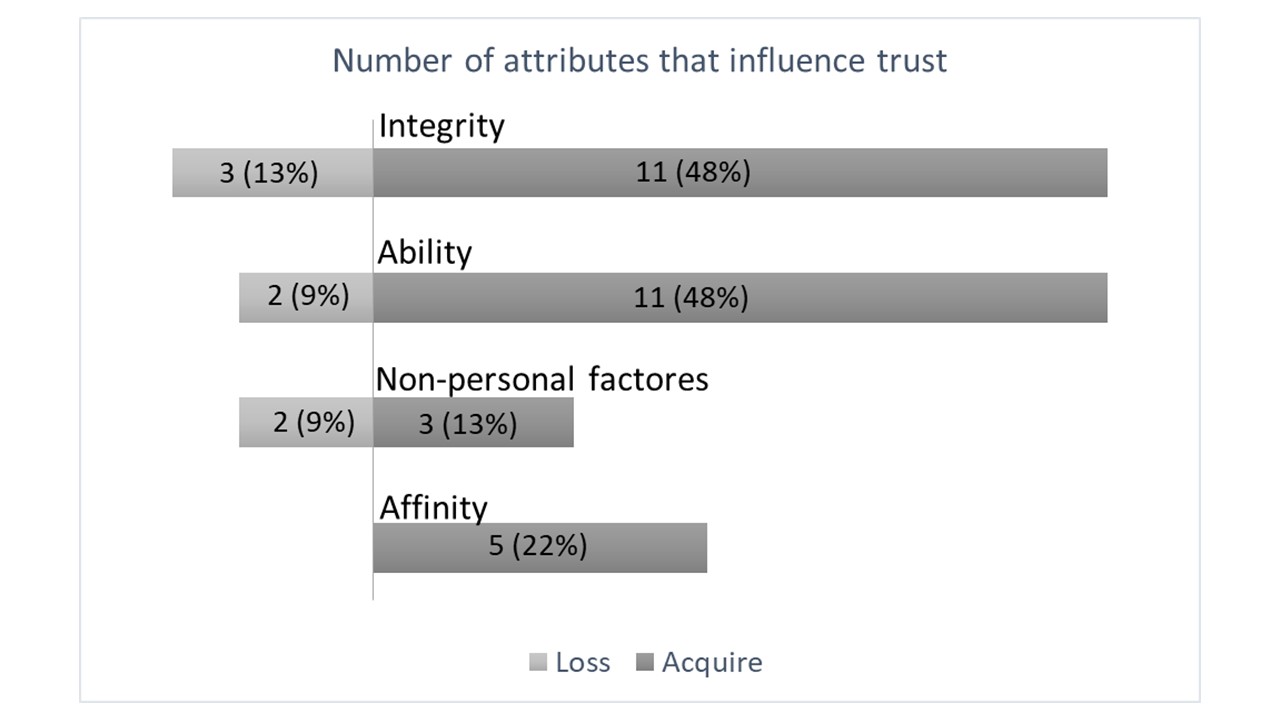}
    \caption{ Attributes themes related to changes in trust phases}
  
    \label{acquire_loss_trust}
\end{figure}

We found that only the phases of acquiring and losing trust were addressed. Specifically, 29 articles considered the  acquiring trust  phase (
\hyperlink{A1}{A1}, \hyperlink{A3}{A3}, \hyperlink{A4}{A4}, \hyperlink{A5}{A5}, \hyperlink{A7}{A7}, \hyperlink{A8}{A8}, \hyperlink{A10}{A10}, \hyperlink{A11}{A11}, \hyperlink{A13}{A13}, \hyperlink{A14}{A14}, \hyperlink{A16}{A16}, \hyperlink{A18}{A18}, \hyperlink{A20}{A20}, \hyperlink{A21}{A21}, \hyperlink{A22}{A22}, \hyperlink{A23}{A23}, \hyperlink{A26}{A26}, \hyperlink{A27}{A27}, \hyperlink{A28}{A28}, \hyperlink{A30}{A30}, \hyperlink{A32}{A32}, \hyperlink{A34}{A34}, \hyperlink{A38}{A38}, \hyperlink{A40}{A40}, \hyperlink{A41}{A41}, \hyperlink{A42}{A42}, \hyperlink{A44}{A44}, \hyperlink{A45}{A45}, \hyperlink{A46}{A46}), while losing trust was reported by 6 articles (\hyperlink{A3}{A3}, \hyperlink{A9}{A9}, \hyperlink{A10}{A10}, \hyperlink{A28}{A28}, \hyperlink{A30}{A30}, \hyperlink{A46}{A46}). Figure \ref{acquire_loss_trust} shows the number of occurrences of attributes that influence the acquiring and losing trust per interpersonal trust themes identified in the literature. The dark grey lines indicate those that cause an acquiring in trust, while those in light grey represent those that lead to a losing of trust. Characteristics that positively affect trust were considered in 63\% (29) of the articles. As we can see, the acquiring trust phase is predominant compared to the losing trust phase. As a highlight, attributes grouped under the AFFINITY theme were only identified in the context of the acquiring phase, which may related to that the willingness to trust could be enhanced by affinity among peers, but its absence might not be a critical factor in losing trust. 

The interpersonal attributes that cause an acquiring trust were classified according to four themes: ABILITY, INTEGRITY, AFFINITY, and NON-PERSONAL FACTORS (Figure \ref {fig:Map}). As stated before, attributes painted in blue are related to the acquiring trust phase, and light grey is related to losing trust. 
Regarding the attributes that can cause trust acquiring, INTEGRITY (12) and ABILITY (11) were the themes with the most attributes. AFFINITY has five attributes, and NON-PERSONAL FACTORS are three. About the losing of trust, we identified three attributes of INTEGRITY and two attributes associated with ABILITY and  NON-PERSONAL FACTORS (See figure \ref{acquire_loss_trust}).

Regarding the ABILITY theme, the categories \textbf{academic skills} (\hyperlink{A20}{A20}, \hyperlink{A46}{A46}), \textbf{collaborative behavior} (\hyperlink{A7}{A7}, \hyperlink{A10}{A10}, \hyperlink{A13}{A13}, \hyperlink{A16}{A16}, \hyperlink{A27}{A27}, \hyperlink{A28}{A28}, \hyperlink{A30}{A30}), \textbf{communication skills} (\hyperlink{A14}{A14}, \hyperlink{A27}{A27}, \hyperlink{A30}{A30}, \hyperlink{A32}{A32}, \hyperlink{A45}{A45}, \hyperlink{A46}{A46}) and \textbf{socio-emotional skills} (\hyperlink{A1}{A1}, \hyperlink{A23}{A23})  can influence the acquiring of trust. However, the \textbf{result-based} category (\hyperlink{A10}{A10}, \hyperlink{A18}{A18}, \hyperlink{A46}{A46}) has attributes that can influence both the acquiring and losing of trust. In this case, the result is the factor analyzed to determine trust. E.g., good results (\textit{high grades}, \hyperlink{A18}{A18}, \hyperlink{A46}{A46}) increase trust, while bad results (\textit{low grades}, \hyperlink{A46}{A46}) reduce trust. Moreover, \textit{hasty decisions} (\hyperlink{A10}{A10})  has the potential to result in unfavorable outcomes.

Concerning to the INTEGRITY theme, the attributes of the categories \textbf{ethical behavior} (\hyperlink{A23}{A23}, \hyperlink{A30}{A30}, \hyperlink{A34}{A34}), \textbf{reputation} (\hyperlink{A11}{A11}, \hyperlink{A26}{A26}), \textbf{transparent behavior} (\hyperlink{A32}{A32}, \hyperlink{A41}{A41}, \hyperlink{A44}{A44}) can influence the acquiring of trust. However,  two categories have attributes that have the power to influence trust to be losing: \textbf{egocentric behavior} (\hyperlink{A10}{A10}), when the student acts motivated only by their own benefit, and \textbf{evasive behavior} (\hyperlink{A10}{A10}), when the student \textit{does not assume error} or \textit{not taking responsibility}. Consequently, the greatest number of attributes within the set of attributes about INTEGRITY is crucial in losing trust.

\begin{figure}
    \centering
    \includegraphics[width=1\linewidth]{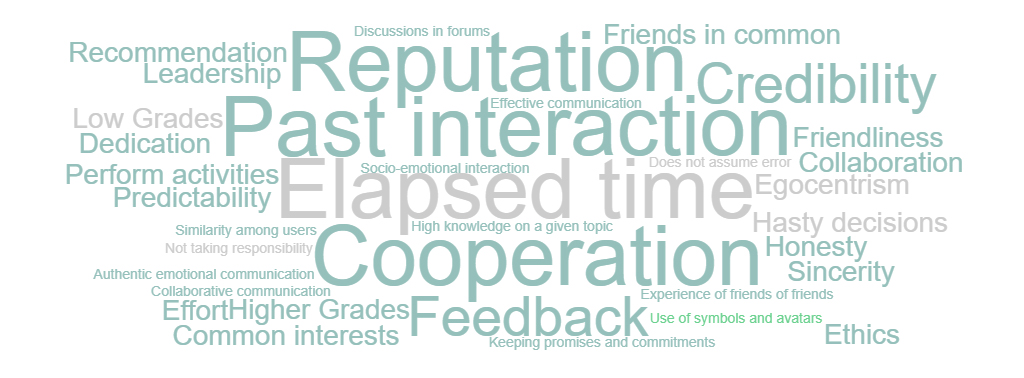}
    \caption{Word cloud with the attributes of interpersonal trust in VLEs}
    \label{fig:word_cloud}
\end{figure}

 In the NON-PERSONAL FACTORS theme, we identified  \textit{elapsed time} (\hyperlink{A9}{A9}, \hyperlink{A28}{A28}, \hyperlink{A30}{A30}) as an attribute that can result in a losing of trust. Trust declines as new evidence of trust does not appear over time. As for the \textbf{non-verbal communication} (\hyperlink{A3}{A3}) category, it has an attribute capable of influencing trust in both its acquiring and losing. The \textit{use of symbols and avatars} is an attribute that, depending on the receptor of the information, the context, symbol or avatar may be interpreted as a concept that either favors or does not trust. In addition, we identified \textbf{second-hand} (\hyperlink{A28}{A28}, \hyperlink{A30}{A30}) trust, which can use a trust network, allowing students to make use of the successful experience of their friends (\textit{experience of friends of his friends}) through \textit{recommendation}. \textbf{Second-hand} trust has no attributes that cause losing of trust because it is a strategy recommendation systems use to recommend trustworthy students based on characteristics that cause acquiring trust. E.g., when a student positively evaluates an interaction with another student, the recommendation system can suggest future interactions based on this evaluation. In this case, when a student receives a recommendation based on a friend's evaluation of another student they interacted with, that recommendation is based on second-hand trust. Figure \ref{fig:word_cloud}  presents the attributes associated with acquiring interpersonal trust among students in light blue and the attributes that can result in losing trust in light gray.

The most frequently occurring words in the word cloud indicate the attributes most frequently identified in the articles. The most frequently addressed topics are \textit{reputation}, \textit{past interaction}, \textit{collaboration}, and \textit{elapsed time}. As a precursor to trust, \textit{reputation} can prompt one student to repose trust in another based on the assessment of trustworthiness derived from these attributes and corroborated by third parties. \textit{Past interactions} represent the most direct and genuine means of assessing an individual's trustworthiness. Based on these experiences, students can evaluate another student's trustworthiness based on their own perceptions. \textit{Collaboration} denotes the joint endeavors undertaken by students in a given activity. E.g., \textit{direct experience} resulting from \textbf{past interactions} represents a form of \textit{collaboration} among students. Additionally, the \textit{elapsed time} is referenced in some articles. The passing of time can potentially erode trust when interactions become less frequent, eventually leading to the dissolution of the relationship and the absence of any remaining evidence of trustworthiness.


As a highlight, few articles address more than one trust phase (\hyperlink{A3}{A3}, \hyperlink{A10}{A10}, \hyperlink{A28}{A28}, \hyperlink{A30}{A30}, \hyperlink{A46}{A46}), which may influence the understanding of the aspects that cause variation in trust, especially concerning the losing of trust. As an exception, \hyperlink{A3}{A3} identified \textit{symbols and avatars} as an attribute that can influence both the acquiring and losing trust (see Figure \ref{fig:Map}).  \hyperlink{A10}{A10} proposed the use of various attributes to compose a student's trustworthiness, including \textit{honest communication, keeps promises and commitments}, and \textit{cooperation} for the trust acquiring phase, while \textit{egocentrism, hasty decisions, does not assume error}, and \textit{not taking responsibility} were attributes considered in trust losing phase. \hyperlink{A28}{A28}, \hyperlink{A30}{A30}, and \hyperlink{A46}{A46} considered \textit{peer recommendation, student reputation, collaboration among students, past interactions, honesty, and feedback in interactions} in the trust acquiring phase. Additionally, \hyperlink{A30}{A30} also considered \textit{similar interests, tastes, and preferences} as trust acquiring attributes, and \hyperlink{A46}{A46} considered \textit{high knowledge on a given topic} and \textit{high grades}. \hyperlink{A46}{A46} regarded \textit{low grades} as a factor in reducing trustworthiness. Meanwhile, both \hyperlink{A28}{A28} and \hyperlink{A30}{A30} considered \textit{elapsed time} as an attribute capable of causing trust reduction or losing. The importance of dynamically updating trust to account for \textit{elapsed time} between interactions is emphasized. Furthermore, a limited number of articles in the literature endeavor to identify the behaviors that can influence users' levels of trust, particularly the behaviors that negatively impact one user's perception of another, leading to a decline in trust, as well as the behaviors that facilitate the maintain and restore of trust. 

In summary, understanding the phases of trust and their associated attributes is crucial for supporting the dynamic aspects of trust. Attributes that cause acquiring of trust can enhance student interactions, while losing trust can reduce collaboration among students in VLEs. Therefore, researching these phases and attributes is essential to developing strategies to support interpersonal trust in the VLE.


\begin{shadedbox}
\textbf{RQ1.3 Key Takeaways:} 
\begin{itemize}

\item  37 interpersonal trust attributes were identified in 32 articles.

\item Acquiring trust is the phase most addressed by the articles, followed by losing trust.

\item Maintain and restore trust phases are not addressed by the articles.

\item Changing trust levels is not an issue among the articles.

\item AFFINITY is only associated with attributes that lead to acquiring trust.

\item Result-based, Egocentric Behavior, Evasive behavior, and Time Categories are associated with the losing the trust phase.

\item Some attributes not directly related to the student's relationship can influence trust, such as \textit{Use of symbols and avatars}, \textit{Experience of friends of his friends}, \textit{Recommendation}, and \textit{Elapsed Time}.

\end{itemize}

\end{shadedbox}

\subsection{\textbf{RQ 2. Which functionalities in VLEs support student interpersonal trust?}}

This section will present the findings regarding the support features for trust among students in VLEs. The data extraction process entailed a detailed observation of the features mentioned in the articles and a comprehensive analysis of their alignment with the attributes and trust phases. The initial step involved listing the identified features, followed by assessing each one to ascertain whether the article is associated with any of the trust attributes. E.g., forums, wikis, and blogs were classified as collaboration tools (collaborative behavior category) because they facilitate knowledge exchange and the collective construction of solutions. Similarly, we analyzed chat rooms and private messaging as practical means of supporting communication, as they enhance communication skills. Additionally, student profiles were examined as tools to support affinity among students by identifying personal, professional, and academic information that can generate affinity. Finally, recommendation systems were assessed to develop second-hand trust by recommending trustworthy peers, even without prior interactions.

\begin{figure}[h]
\includegraphics[width=.9\textwidth]{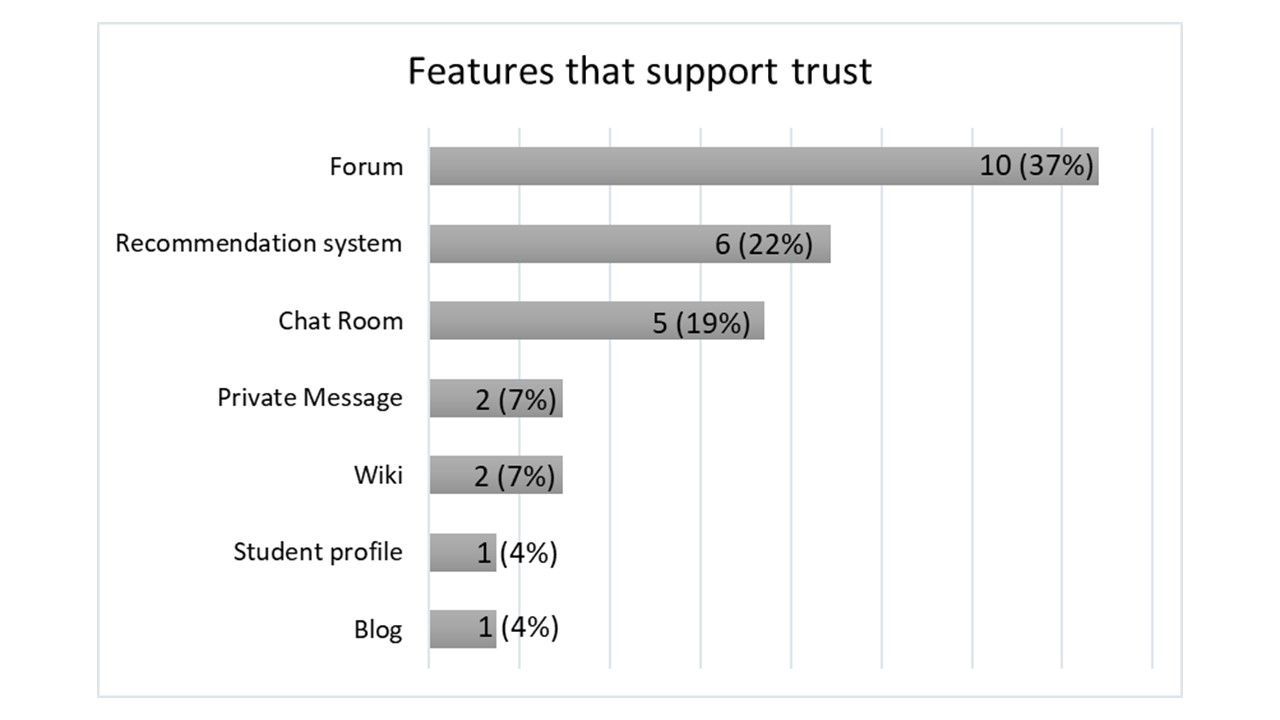}
\caption{Features in VLEs that support trust among students}
\label{fig:Features in E-learning}
\end{figure}

Figure \ref{fig:Features in E-learning} illustrates the extracted features, their frequencies, and percentages about the total number of identified features. Some features, such as chat rooms, blogs, wikis, and private messages, indirectly facilitate trust formation as they allow students to communicate and collaborate, thereby facilitating the extraction of data that influences trust. In terms of directly supporting interpersonal trust, forums provide \textit{feedback}  to students by helping to characterize responses. Positively evaluated answers acquiring trust and indicate a \textit{high knowledge on a given topic}. In addition, students' active \textit{discussions in forums} contribute to acquiring trust based on the student's \textbf{collaborative behavior}.

The student profile is a feature that enables students to identify other students based on their personal and professional characteristics, as outlined in the profile, by allowing students to develop trust through the identification \textbf{similarity among students}. Furthermore, the profile can be updated dynamically with student information of \textit{collaboration}, including data on \textit{perform activities}  and academic performance (\textbf{result-based}).

Another feature that directly supports interpersonal trust in VLEs is recommendation systems. Recommendation systems mainly use the \textbf{similarity among students} to recommend trustworthy students for \textit{collaboration}. They use trust networks by spreading trust based on  \textit{friends in common} that students have, spreading \textit{experience of friends of his friends} that have generated trust, as well as exploiting \textbf{similarity among students} data such as \textit{friends in common} in social networks.

Regarding features in the selected articles, \textit{Collaboration} and \textit{communication} features, such as forums, were identified in 10 articles (37\%)  (\hyperlink{A1}{A1}, \hyperlink{A7}{A7}, \hyperlink{A9}{A9}, \hyperlink{A10}{A10}, \hyperlink{A14}{A14}, \hyperlink{A17}{A17}, \hyperlink{A25}{A25}, \hyperlink{A27}{A27}, \hyperlink{A36}{A36}), \textit{wikis} in 2 articles (7\%) (\hyperlink{A23}{A23}, \hyperlink{A27}{A27}) and \textit{blogs} in 1 article (4\%) (\hyperlink{A23}{A23}), \textit{chat} in 5 articles (19\%) (\hyperlink{A3}{A3}, \hyperlink{A17}{A17}, \hyperlink{A25}{A25}, \hyperlink{A32}{A32}, \hyperlink{A36}{A36}) and \textit{private messaging} in 2 articles (7\%) (\hyperlink{A17}{A17},  \hyperlink{A36}{A36}) were cited as supporting collaboration among students in VLEs, as features that indirectly support trust. 

Unlike the articles mentioned above, the following articles provide specific features for trust support. \hyperlink{A15}{A15} proposes measured student participation in forums to predict students' trust values, using  \textit{reputation} and \textit{direct experience} among students to support trust. \textit{Direct experience} is calculated from user  (\textbf{past interactions}) in activities and forums. \textit{Reputation} is calculated based on \textit{direct experience} between students, derived from the history of student opinions about others, expressed through positive or negative interactions, such as the use of "liked" or "disliked" icons obtained on forums or other interaction tools.

Recommendation systems are the most used specific strategy to support interpersonal trust. (\hyperlink{A30}{A30}, \hyperlink{A40}{A40}, \hyperlink{A22}{A22}, \hyperlink{A41}{A41}, \hyperlink{A4}{A4}). Usually, this strategy allows for initial trust based on \textbf{similarities among users} or according to some attribute they have in common, such as \textit{friends in common}, or based on the successful experience of others (\textit{experience of friends of his friends}) to recommend a trustworthy student in a trust network. 

\hyperlink{A30}{A30} presented a dynamic peer recommendation strategy for MOOC courses, proposing a recommendation system architecture that aids in selecting trustworthy partners for collaboration. The system considers several attributes: \textit{elapsed time} since the last interaction, as trust can decrease over time; \textit{feedback} from interactions to gauge trustworthiness; social \textit{cooperation} between students to determine collaboration potential; and \textit{similar interests, tastes, and preferences} obtained from online social networks to assess \textbf{similarity among users}. Additionally, \textit{honesty} is evaluated as a social factor alongside interaction assessments. The recommendation can also utilize \textbf{second-hand} trust, leveraging the \textit{experiences of friends of friends}. Despite addressing multiple attributes in its proposal, \hyperlink{A30}{A30} did not evaluate the system in a real-world scenario.

\hyperlink{A40}{A40} to use a recommendation strategy based on matrices and tags to characterize the students and derive similarity values.   The trust matrix contains the explicit social relationships among students collected from social networks. The tags contain the student's classification of items within the system, such as evaluating a forum. Finally, the \textbf{similarity among students} was calculated based on the characteristics of the students contained in the relationship matrices for each student, as well as the similarity of the values of the students' tags, i.e., the similar evaluations that the students have made.

\hyperlink{A22}{A22} uses a collaborative filtering strategy based on deep neural networks that learn student preferences. It is able to deeply explore each student's social network interactions, and the trust-aware attention layer is designed to indicate each student's social influence.  It facilitates the spread of trust in a network of students, allowing students to benefit from the trust information shared by other classmates who have common connections, i.e., \textit{friends in common}.

\hyperlink{A41}{A41} proposed a learning recommendation method that uses collaborative filtering based on social student trust. To do this, it incorporates measures of trust among students to improve the accuracy of recommendations. This involves assessing trustworthiness based on \textbf{past interactions} among students. The degree of trust is influenced by the quality of interactions, with successful interactions increasing trust. Trust is estimated on the basis of these interactions, and a final calculation takes student preferences into account, resulting in the students' trust network after removing relationships with low trust.

\hyperlink{A4}{A4} used trust-based recommendations for e-learning to facilitate \textit{collaboration} between students by recommending trustworthy and relevant study partners, taking into account students' learning styles and knowledge levels, aiming to integrate trust measures to improve the accuracy of recommendations. This involves assessing the trust between students based on the similarity of the features that the students have evaluated, where the common features that they have evaluated indicate the \textbf{similarity among students}.

Several attributes identified in our analysis were not related to the features discussed in the articles. Most of them, such as \textit{Authentic emotional communication} (\hyperlink{A44}{A44}), \textit{Collaborative communication} (\hyperlink{A27}{A27}), \textit{Cooperation} (\hyperlink{A13}{A13}), \textit{Credibility} (\hyperlink{A11}{A11}, \hyperlink{A26}{A26}), \textit{Dedication} (\hyperlink{A32}{A32}), \textit{Does not assume error} (\hyperlink{A10}{A10}), \textit{Effort} (\hyperlink{A32}{A32}), \textit{Egocentrism} (\hyperlink{A10}{A10}), \textit{Ethics} (\hyperlink{A23}{A23}), \textit{Hasty decisions} (\hyperlink{A10}{A10}), \textit{Honest communication} (\hyperlink{A10}{A10}), \textit{Honesty} (\hyperlink{A30}{A30}, \hyperlink{A34}{A34}), \textit{Keeping promises and commitments} (\hyperlink{A10}{A10}), \textit{Leadership} (\hyperlink{A20}{A20}), \textit{Not taking responsibility} (\hyperlink{A10}{A10}), \textit{Predictability} (\hyperlink{A41}{A41}), \textit{Sincerity} (\hyperlink{A32}{A32}), \textit{Socio-emotional interaction} (\hyperlink{A23}{A23}) were cited in the articles as capable of influencing trust, but were not related in their proposals or analyses to any feature of VLEs. In addition to these attributes, some others, such as \textit{Effective communication} (\hyperlink{A32}{A32}), \textit{Friendliness} (\hyperlink{A1}{A1}), \textit{Higher Grades} (\hyperlink{A46}{A46}), \textit{Low Grades} (\hyperlink{A46}{A46}), \textit{Use of symbols and avatars} (\hyperlink{A3}{A3}) were also not related to features, as they were findings performed in the articles.

In summary, the use of recommendation systems and student trustworthiness profiles are related to the support and development of initial trust in VLEs. According to \cite{koufaris2004development}, initial trust is the initial belief of trust that can be formed without any previous experience or interaction between two parties. In this case, the recommendation of a trustworthy student by a system developed for this purpose or the perception of a student's personal and professional characteristics in an identification panel can influence the acquiring of trust in VLEs. In this sense, the attributes of trust that influence both the acquiring phase and the losing of trust are used to define whether a student is trustworthy to another.
    
\begin{shadedbox}
 \textbf{RQ 2. Key Takeaways:} 

 \begin{itemize}
     \item Communication and collaboration features are used most to support interpersonal trust in VLEs. 
 
    \item User profiles emerge as the functionalities that support acquiring trust and initial trust.

    \item Recommendation systems are used to identify trustworthy students who are not directly part of the student's trust network.

    \item Forum, chat, and private messaging are valuable sources of data to be exploited to build student trustworthiness.

    \item Data mining and neural networks are approaches used to identify attributes that influence trust within a network.

    \item Some attributes that are conducive to fostering trust are not being considered in the proposals for VLE features. 
    
    \item Most features presented have not been tested in a real-world scenario, but some recommendation systems have tested their approach on public databases.
\end{itemize}
 
\end{shadedbox}


\section{Discussion}

 Since real-world activities are supported by computers, trust is a crucial aspect to be considered in interpersonal relationships. Therefore, trust has been studied in several research fields such as sociology \cite{sztompka1999trust}, psychology \citep{robinson1996trust}, economics \citep{knack2003building, berggren2006free}, communication \citep{cheung2013interweaving}, leadership \citep{burke2007trust}, and the implementation of self-managed work teams \citep{trainer2018bridging}. More recently, due to the popularization of distance and remote learning environments and tools, trust aspects are studied under this domain, as they are fundamental to success in the teaching-learning process,  influencing not only students' interactions but also the overall effectiveness of collaborative learning, communication, and educational outcomes, helping building a supportive learning community \citep{wang_2014_building_student_trust}. Understanding the intricacies of interpersonal trust within VLEs can provide insights into enhancing these educational platforms to better serve students and educators. Hence, this SMS offers a comprehensive view of various attributes, phases, and features addressed in the current literature.

In light of the analysis of the definitions and the principal terms used by the selected articles (see Table \ref{tab:table_def}), based on \cite{mayer1995integrative} and our main findings, we put forth the following definition of interpersonal trust for virtual learning environments: \textit{``Trust is the expectation that leads students to interact and collaborate in virtual learning environments to perform academic activities. Students trustor select their partners based on affinities or personal assessments of the student trustee's integrity, abilities, and likelihood of acting under expectations in a given context, which may result in positive academic outcomes. Once trust is established, it can be built, maintained, lost, and rebuilt over time."}

The widely used \cite{mayer1995integrative} attributes classification has been extended in this SMS by affinity and non-personal factors. AFFINITY plays a fundamental role in building interpersonal trust; it can manifest itself in various ways within a VLE, mostly based on \textit{direct experience} among students, common educational goals and \textit{friends in common} \citep{elghomary_2019_dynamic_peer, chamba_tvlc2023, Elghomary_Bouzidi_Daoudi_2022_design_of_smart_MOOC}.  In addition, AFFINITY can facilitate \textit{communication} and \textit{collaboration} among students. When students feel connected to each other, they are more likely to share ideas, ask for help, and work together on academic projects. This fosters an environment of mutual support and \textit{cooperation}, further strengthening the bonds of trust among VLE participants. These personal connections help to create a sense of community and belonging, which are fundamental for establishing trusting relationships.

Understanding attributes is essential for supporting the dynamic aspects of trust. As identified, a vast list of attributes is used to assess interpersonal trust among students in VLEs (About 37 attributes, see Figure \ref{fig:Map}). The list of attributes is extensive, underscoring the fact that the perception of trust is highly personal and subjective, depending on the personal perspective of the trustor (e.g., cultural and social aspects), may varing significantly from one trustor to another, and are difficult to associate with quantifiable indicators \citep{wang2005overview}. The challenge of trust in the virtual world is significant compared to the real world, as in the physical world, numerous elements can be perceived and evaluated, making it relatively simpler to assess trust compared to virtual environments. In a virtual environment, the absence of tangible physical signs and the reliance on digital interactions complicates the assessment of trust, rendering it a more complex and nuanced process \citep{anwar_2021_supporting_privacy}. Distinct of others domains, e.g., business domain,  trust for academic collaboration involves more nuanced and sensitive attributes, reflecting individual characteristics and varying perceptions about the appropriate knowledge, skills and behavioral tendencies to develop relationships \citep{schaubroeck2013developing}. These attributes are inherently less quantifiable, presenting a greater challenge for support in virtual environments. However, given the pervasive role of software in modern society, addressing these challenges is imperative. Current research efforts are devoted to providing theoretical models and developing features to support the various aspects of trust relationships \citep{A2, wang_2014_building_student_trust, udhayakumar_2021_trusted_cooperative, valtolina_2022_extended_utaut}.

Understanding the trust phases and their associated attributes is crucial for supporting the dynamic aspects of trust in VLEs. Attributes that enhance trust can significantly amplify \textit{collaboration} among students, while those that erode trust can diminish collaborative efforts. With regard to the evolutionary phases of trust and initial trust,  acquiring phase is widely explored in the selected  articles, demonstrating a concern with the need to build trust in VLEs. However, the maintain and restoration phases were not explored, despite its importance in establishing long term trusting relationships \citep{fachrunnisa2010state}. Maintain trust in the context of the brevity of online courses may not have been prioritized due to the short timeframe, relative to the fast-paced nature of online courses \cite{anwar_2021_supporting_privacy}. As for the losing of trust, some attributes were identified, but the literature still lacks knowledge about the aspects that can erode trust. Beyond negative behavior, the \textit{elapsed time} is highly associated, which requires trust to be updated dynamically due to natural oscillations that consequently cause a losing of trust \citep{elghomary_2019_dynamic_peer}. Therefore, researching these phases and attributes is essential to develop strategies that enhance trust, mitigate its erosion, and ultimately foster a more collaborative and effective virtual learning environment. As we continue to integrate technology into education, it becomes increasingly important to address these complexities in order to create VLEs that foster trust among students. \citep{zhang_2022_a_novel_precise}.

The dynamics of trust have been fostered by considering that trust will vary over time according to events that may impact it.   The trust assessment needs to be updated, and the trust status needs to be constantly evolving \citep{elghomary_2019_dynamic_peer}. In this sense, the phases of trust must be considered so that the updating process is effective and the student can have their trust status updated positively or negatively. Therefore, specific strategies for maintain trust over time may be needed to ensure more effective collaboration processes in VLEs.

With regard to strategies to support trust in the VLE, recommendation systems, and user profiles were the main support for the aspects of trust mentioned above. Overall, these strategies support the transitivity (\textbf{Second-hand} trust) and dynamic trust properties.  Recommendation systems are being used to identify trustworthy students, encourage knowledge sharing, and collaborate in academic activities,  promoting trust networks to connect students by their affinities \citep{kunthi_2018_exploring_antecedent, elghomary_2019_dynamic_peer}. Moreover trust can be spread within the network of friends through trust propagation, with recommendation systems propagating a student's trust to other students within the network \citep{mcevily2021network}. In addition, trust networks allow the transitivity property of trust to be explored on the basis of \textit{friends in common}. Data from interaction in communication tools, collaboration tools and the evaluation of \textit{cooperation} among students can be used to feed recommendation systems. 
 The dynamics of trust were supported by updating student trust profiles, considering that student behavior is dynamic and evolves during the learning process \citep{miguel2014collective}. Identifying trustworthy colleagues through the use of trust profiles can help highlight the characteristics that inspire trust in other students. The dynamic property of trust is taken into account by keeping the trust profile updated as actions that influence trust occur. A dynamically updating trust architecture is based on continuously changing data inputs and periodically updates users' trust calculations, which means that the trust value can change along with interactions. As highlighted,  the asymmetric property of trust should be considered, as it indicates that one student can trust another without reciprocity; in this context, interactions between students that generate trust data should be evaluated by both students involved \citep{golbeck2005computing}. The evaluation of \textit{cooperation} makes use of the asymmetric property of trust, allowing students to evaluate their satisfaction in cooperative interactions \citep{chan_2016_trust_ware_peer,elghomary_2019_dynamic_peer}.

Investigate new characteristics that potentially boost trust should promote a more favorable environment for learning. Therefore, based on the SMS findings, we can non-exhaustively suggest several research opportunities:

\begin{itemize}
     \item \textbf{Interpersonal trust from the students perspective}. Although several attributes are addressed by the literature, little consideration was given to the student's perspective on the factors that cause trust to change having the power to influence interpersonal trust. With the significant growth in the use of online courses, it is essential to adopt qualitative approaches, such as surveys, interviews, and case studies, to explore the characteristics that influence the evolutionary phases of interpersonal. In addition, the use of comparative analysis in different contexts can identify which attributes are most effective in restore and maintain trust. This provides a broader understanding of the factors influencing trust in these evolutionary phases and whether attributes from other contexts can be applied to online courses. A student-centered approach should reveal news 
 aspects, enhance trust, and consequently, learning effectiveness.  
     
 \item \textbf{Interpersonal trust from the educational professional perspective}. Education professionals can foster the establishment of trust among students by using pedagogical strategies focused on collaboration among students aimed at the collaborative construction of content, discussion of content in forums and chats, peer evaluation, and the sharing of content and knowledge in VLEs. Similarly, from the student's point of view, adopting qualitative approaches, such as surveys, interviews, and case studies, to evaluate teaching strategies and activities that influence trust in order to optimize teaching strategies may bring interesting findings. 

\item \textbf{Trust Continuous Evaluation}.
At present, research on interpersonal trust between students in virtual learning environments (VLEs) is largely isolated, with studies focusing on specific instances without considering the long-term or continuous nature of trust. To gain insight into the attributes of trust in online learning environments, it is essential to conduct longitudinal studies that can elucidate the factors influencing the phases of trust. Given that students often enroll in multiple courses on the same learning platform, such studies are crucial to understanding the re-establishment of trust and the recovery process after damaging events.  As the information on the trust network is observed in several courses, it can be used to influence the students' performance, retention, and success in academic activities.

 \item \textbf{Beyond Initial Trust} Understanding the factors that influence the different evolutionary phases of trust is fundamental to developing strategies aimed at interpersonal trust in VLEs. This makes it possible to create environments that promote the acquiring and maintain of trust, as well as dealing effectively with situations that can damage it, including the need for restoration after events that cause it to weaken. Identify factors that cause losing of trust in virtual educational contexts and understand their implications; Analyze how trust can be maintain among students in online learning environments in order to promote stronger relationships;
 To study which behaviors and attitudes can restore trust among students, with a view to rebuilding damaged relationships.
 As a highlight, the articles reveal a limited focus on evaluating trust attributes during a course.
     \item \textbf{Personalized Trust}. 
Trust is subjective and the assessment is very personal. Use of profiles with data on student characteristics that allow the development of trust based on personal and professional affinity, as well as a trust score in different attribute that allows students to identify, based on their needs, which colleagues can best meet them during collaborative activities. Moreover, investigate the transparency of the recommendations provided by recommendation systems, making clear the criteria used for the recommendation, since the characteristics that one student considers essential for trusting another may not be the same for another student, and the final decision as to whether a student is trustworthy for collaboration should be centered on the student;

\item \textbf{Emerging Technologies and Trust}. The advent of novel technologies, such as artificial intelligence (AI) and machine learning (ML), offers substantial potential for fostering trust within VLEs. Such technologies may be utilized to oversee interactions between students and to identify behaviors that impact trust. By analyzing patterns and trends in student interactions, artificial intelligence (AI) and machine learning (ML) can identify key indicators of trustworthiness, such as responsiveness, collaboration frequency, and the quality of contributions. Moreover, these technologies can be utilized to construct intricate reputation systems that consolidate disparate trust indicators into a unified trust score for each student. It is of particular importance to adopt a privacy-by-design approach when incorporating AI and ML technologies into VLEs. This necessitates the integration of data protection principles into the design and development of AI systems from the outset. Regulations such as the General Data Protection Regulation (GDPR) in Europe and the General Data Protection Law (Lei Geral de Proteção de Dados - LGPD) in Brazil impose rigorous requirements for data collection, processing, and storage. These laws underscore the necessity for transparency, consent, and the right to privacy for individuals whose data is being collected.

 \item \textbf{Trust Body of Knowleddge}. A Trust Body of Knowledge (TrustBOK) for virtual learning environments (VLEs) may offers a solid theoretical foundation for proposals and reports, comprehensive, theoretically sound, and practically applicable understanding of trust concept, ensuring that they are grounded in established concepts. The TrustBOK should comprise conceptual frameworks, models, and metamodels. While conceptual frameworks offer a structured approach to understanding the multifaceted nature of trust, the use of metamodels and models aim to provide higher-level abstractions of trust relationships can facilitate the design and implementation of trust-enhancing features in VLEs. TrustBOK can provide clear guidance for industry and academia to enhance trust in creating more robust and trustworthy educational platforms.

\end{itemize}


\subsection{Threats to validity}

We considered the limitations presented by \cite{kitchenham2015evidence} as threats to the validity of our study, where she discusses the main limitations in secondary studies. 

\textbf{Construct validity} reflects the extent to which the operational measures studied represent the concept the researcher intends to investigate, as defined by the research questions. This concern encompasses the capability of the study design to effectively address the research question \citep{kitchenham2015evidence, wohlin2012experimentation}. To minimize the risks of construct validity, we adopted a rigorous process in developing the  SMS research protocol. Two researchers drew up this protocol, went through several stages of revision and refinement, and was validated by a third researcher to ensure impartiality and accuracy. To analyze the instrument's validity, we performed a pilot data extraction. This procedure aimed to assess whether the answers to the research questions were satisfactory and to adjust any inconsistencies found. The pilot phase made it possible to identify and correct potential flaws in the measuring instrument, thus ensuring that it adequately captured the intended concepts.

\textbf{Internal validity} concern is with the conduct of the study, mainly related to data extraction and synthesis, and whether there are factors that may have caused some degree of bias in the overall process \citep{kitchenham2015evidence}. To mitigate threats to internal validity, the study was conducted by two researchers and reviewed by a third member of our research group. However, for an SMS, a lower level of comprehensiveness of the primary studies returned may be acceptable \citep{kitchenham2015evidence}. Two researchers then independently extracted data according to the RQs we wanted to answer. Subsequently, data synthesis was performed jointly by the two researchers. A third researcher validated the whole process.

\textbf{External validity} for a secondary study should be based on assessing the comprehensiveness of the primary studies.  Three databases were chosen (IEEE, ACM, and Scopus), as they comprise the most important journals and conferences in software engineering and computing, in general, to minimize this risk \citep{kitchenham2015evidence}.  To test the effectiveness of the search string, we used control studies that we knew should be returned in each database. We calibrated the string until all the studies were returned.  Additionally, two researchers applied the inclusion/exclusion criteria separately. A stricter analysis could have ruled out articles relevant to this study. Furthermore, the third researcher carried out a sample check on 10\% of the excluded articles to check that the exclusion criteria had been applied correctly. 
Despite this, \cite{kitchenham2015evidence} states that in secondary studies that aim to identify motivating factors, which is the case in our study, where common factors are present in several studies, it does not substantially affect the result.  


\section{Final considerations and future work}
\normalsize

 This SMS focused on interpersonal trust among students in virtual learning environments. Regarding this type of trust, there is a need for further exploration of the factors that can influence it. Identifying and categorizing these attributes provides a solid basis for developing tools and systems that promote trust and collaboration. Understanding trust dynamics may offer promising avenues for improving interactions and academic performance in VLE.
In summary, the literature considers several attributes, phases, and features that support interpersonal trust among students in VLEs. However, we identified gaps in understanding the behavioral characteristics of students that affect trust in VLEs. Moreover,  many studies addressed trust acquiring, but trust losing was under-explored. Additionally, the literature mainly focuses on acquiring trust,  requiring yet investigating attributes or solutions to maintain and restore trust.

It is essential to collect data that enables the creation of environments conducive to trust among students in the context of virtual learning environments. When building models or systems based on trust, it is essential to make clear the attributes considered for achieving a specific objective. When building new models, it is essential to consult the literature beforehand so that the models consider requirements that increase and maintain trust, prevent losing, and restore trust when necessary. 

In future work, we intend to identify more attributes influencing interpersonal trust in VLEs. We intend to investigate with VLE students which personal characteristics affect t trust evolutionary phases. In addition, we intend to develop a conceptual model to support trust among virtual students.

\bibliographystyle{elsarticle-harv} 
\bibliography{Manuscript}

\begin{thebibliography}{89}
\expandafter\ifx\csname natexlab\endcsname\relax\def\natexlab#1{#1}\fi
\providecommand{\url}[1]{\texttt{#1}}
\providecommand{\href}[2]{#2}
\providecommand{\path}[1]{#1}
\providecommand{\DOIprefix}{doi:}
\providecommand{\ArXivprefix}{arXiv:}
\providecommand{\URLprefix}{URL: }
\providecommand{\Pubmedprefix}{pmid:}
\providecommand{\doi}[1]{\href{http://dx.doi.org/#1}{\path{#1}}}
\providecommand{\Pubmed}[1]{\href{pmid:#1}{\path{#1}}}
\providecommand{\bibinfo}[2]{#2}
\ifx\xfnm\relax \def\xfnm[#1]{\unskip,\space#1}\fi
\bibitem[{Ahmadian et~al.(2022)Ahmadian, Ahmadian and Jalili}]{ahmadian_2022_a_deep_learning}
\bibinfo{author}{Ahmadian, S.}, \bibinfo{author}{Ahmadian, M.}, \bibinfo{author}{Jalili, M.}, \bibinfo{year}{2022}.
\newblock \bibinfo{title}{A deep learning based trust- and tag-aware recommender system}.
\newblock \bibinfo{journal}{Neurocomputing} \bibinfo{volume}{488}, \bibinfo{pages}{557--571}.
\newblock \DOIprefix\doi{https://doi.org/10.1016/j.neucom.2021.11.064}.
\bibitem[{Anwar(2021)}]{anwar_2021_supporting_privacy}
\bibinfo{author}{Anwar, M.}, \bibinfo{year}{2021}.
\newblock \bibinfo{title}{Supporting privacy, trust, and personalization in online learning}.
\newblock \bibinfo{journal}{International Journal of Artificial Intelligence in Education} \bibinfo{volume}{31}, \bibinfo{pages}{769--783}.
\newblock \DOIprefix\doi{10.1007/s40593-020-00216-0}.
\bibitem[{Arnado et~al.(2021)Arnado, Jabal, Poa and Viray}]{arnado_2021_trust_in_pandemic}
\bibinfo{author}{Arnado, J.M.}, \bibinfo{author}{Jabal, R.F.}, \bibinfo{author}{Poa, M.R.J.A.}, \bibinfo{author}{Viray, T.C.}, \bibinfo{year}{2021}.
\newblock \bibinfo{title}{Trust in pandemic-induced online learning: Competitive advantage of closure and reputation}.
\newblock \bibinfo{journal}{RISE} \bibinfo{volume}{10}, \bibinfo{pages}{192--217}.
\newblock \DOIprefix\doi{10.17583/rise.2021.7088}.
\bibitem[{Bachmann(2001)}]{bachmann2001trust}
\bibinfo{author}{Bachmann, R.}, \bibinfo{year}{2001}.
\newblock \bibinfo{title}{Trust, power and control in trans-organizational relations}.
\newblock \bibinfo{journal}{Organization studies} \bibinfo{volume}{22}, \bibinfo{pages}{337--365}.
\newblock \DOIprefix\doi{https://doi.org/10.1177/0170840601222007}.
\bibitem[{Baturay and Toker(2019)}]{baturay_2019_the_comparison_of}
\bibinfo{author}{Baturay, M.H.}, \bibinfo{author}{Toker, S.}, \bibinfo{year}{2019}.
\newblock \bibinfo{title}{The comparison of trust in virtual and face-to-face collaborative learning teams}.
\newblock \bibinfo{journal}{Turkish Online Journal of Distance Education} \bibinfo{volume}{20}, \bibinfo{pages}{153--164}.
\newblock \DOIprefix\doi{10.17718/tojde.601929}.
\bibitem[{Berggren and Jordahl(2006)}]{berggren2006free}
\bibinfo{author}{Berggren, N.}, \bibinfo{author}{Jordahl, H.}, \bibinfo{year}{2006}.
\newblock \bibinfo{title}{Free to trust: Economic freedom and social capital}.
\newblock \bibinfo{journal}{Kyklos} \bibinfo{volume}{59}, \bibinfo{pages}{141--169}.
\newblock \DOIprefix\doi{https://doi.org/10.1111/j.1467-6435.2006.00324.x}.
\bibitem[{Bhaskaran and Santhi(2019)}]{bhaskaran2019efficient}
\bibinfo{author}{Bhaskaran, S.}, \bibinfo{author}{Santhi, B.}, \bibinfo{year}{2019}.
\newblock \bibinfo{title}{An efficient personalized trust based hybrid recommendation (tbhr) strategy for e-learning system in cloud computing}.
\newblock \bibinfo{journal}{Cluster Computing} \bibinfo{volume}{22}, \bibinfo{pages}{1137--1149}.
\newblock \DOIprefix\doi{https://doi.org/10.1007/s10586-017-1160-5}.
\bibitem[{B{\o}e(2018)}]{boe_2018_learning_technology}
\bibinfo{author}{B{\o}e, T.}, \bibinfo{year}{2018}.
\newblock \bibinfo{title}{E-learning technology and higher education: the impact of organizational trust}.
\newblock \bibinfo{journal}{Tertiary Education and Management} \bibinfo{volume}{24}, \bibinfo{pages}{362--376}.
\newblock \DOIprefix\doi{https://doi.org/10.1080/13583883.2018.1465991}.
\bibitem[{Borum(2010)}]{borum2010science}
\bibinfo{author}{Borum, R.}, \bibinfo{year}{2010}.
\newblock \bibinfo{title}{The science of interpersonal trust}.
\newblock \bibinfo{journal}{Randy Borum} , \bibinfo{pages}{1--79}\URLprefix \url{http://works.bepress.com/randy_borum/48/}.
\bibitem[{Burke et~al.(2007)Burke, Sims, Lazzara and Salas}]{burke2007trust}
\bibinfo{author}{Burke, C.S.}, \bibinfo{author}{Sims, D.E.}, \bibinfo{author}{Lazzara, E.H.}, \bibinfo{author}{Salas, E.}, \bibinfo{year}{2007}.
\newblock \bibinfo{title}{Trust in leadership: A multi-level review and integration}.
\newblock \bibinfo{journal}{The leadership quarterly} \bibinfo{volume}{18}, \bibinfo{pages}{606--632}.
\newblock \DOIprefix\doi{https://doi.org/10.1016/j.leaqua.2007.09.006}.
\bibitem[{Chamba-Eras et~al.(2016)Chamba-Eras, Arruarte and Elorriaga}]{chamba_2016_bayesian_networks}
\bibinfo{author}{Chamba-Eras, L.}, \bibinfo{author}{Arruarte, A.}, \bibinfo{author}{Elorriaga, J.A.}, \bibinfo{year}{2016}.
\newblock \bibinfo{title}{Bayesian networks to predict reputation in virtual learning communities}, in: \bibinfo{booktitle}{2016 IEEE Latin American Conference on Computational Intelligence (LA-CCI)}, \bibinfo{organization}{IEEE}. pp. \bibinfo{pages}{1--6}.
\newblock \DOIprefix\doi{10.1109/LA-CCI.2016.7885721}.
\bibitem[{Chamba-Eras et~al.(2023)Chamba-Eras, Arruarte and Elorriaga}]{chamba_tvlc2023}
\bibinfo{author}{Chamba-Eras, L.}, \bibinfo{author}{Arruarte, A.}, \bibinfo{author}{Elorriaga, J.A.}, \bibinfo{year}{2023}.
\newblock \bibinfo{title}{T-vlc: A trust model for virtual learning communities}.
\newblock \bibinfo{journal}{IEEE Transactions on Learning Technologies} \bibinfo{volume}{16}, \bibinfo{pages}{847--860}.
\newblock \DOIprefix\doi{10.1109/TLT.2023.3288363}.
\bibitem[{Chan et~al.(2016)Chan, Zhao and King}]{chan_2016_trust_ware_peer}
\bibinfo{author}{Chan, H.P.}, \bibinfo{author}{Zhao, T.}, \bibinfo{author}{King, I.}, \bibinfo{year}{2016}.
\newblock \bibinfo{title}{Trust-aware peer assessment using multi-armed bandit algorithms}, in: \bibinfo{booktitle}{Proceedings of the 25th International Conference Companion on World Wide Web}, pp. \bibinfo{pages}{899--903}.
\newblock \DOIprefix\doi{https://doi.org/10.1145/2872518.2891080}.
\bibitem[{Chang(2017)}]{chang_2017_online_training}
\bibinfo{author}{Chang, W.L.}, \bibinfo{year}{2017}.
\newblock \bibinfo{title}{Online training for business plan writing through the world caf{\'e} method: the roles of leadership and trust}.
\newblock \bibinfo{journal}{Universal Access in the Information Society} \bibinfo{volume}{16}, \bibinfo{pages}{313--324}.
\newblock \DOIprefix\doi{https://doi.org/10.1007/s10209-016-0459-y}.
\bibitem[{Cheung et~al.(2013)Cheung, Yiu and Lam}]{cheung2013interweaving}
\bibinfo{author}{Cheung, S.O.}, \bibinfo{author}{Yiu, T.W.}, \bibinfo{author}{Lam, M.C.}, \bibinfo{year}{2013}.
\newblock \bibinfo{title}{Interweaving trust and communication with project performance}.
\newblock \bibinfo{journal}{Journal of construction engineering and management} \bibinfo{volume}{139}, \bibinfo{pages}{941--950}.
\newblock \DOIprefix\doi{https://doi.org/10.1061/(ASCE)CO.1943-7862.0000681}.
\bibitem[{Choi and Cho(2019)}]{choi2019mechanism}
\bibinfo{author}{Choi, O.K.}, \bibinfo{author}{Cho, E.}, \bibinfo{year}{2019}.
\newblock \bibinfo{title}{The mechanism of trust affecting collaboration in virtual teams and the moderating roles of the culture of autonomy and task complexity}.
\newblock \bibinfo{journal}{Computers in Human Behavior} \bibinfo{volume}{91}, \bibinfo{pages}{305--315}.
\newblock \DOIprefix\doi{https://doi.org/10.1016/j.chb.2018.09.032}.
\bibitem[{Clarke and Braun(2013)}]{clarke2013successful}
\bibinfo{author}{Clarke, V.}, \bibinfo{author}{Braun, V.}, \bibinfo{year}{2013}.
\newblock \bibinfo{title}{Successful qualitative research: A practical guide for beginners}.
\newblock \bibinfo{publisher}{Sage publications ltd}.
\newblock \DOIprefix\doi{https://uk.sagepub.com/en-gb/eur/successful-qualitative-research/book233059}.
\bibitem[{Costello et~al.(2018)Costello, Brunton, Brown and Daly}]{costello_2018_In_moocs_we_trust}
\bibinfo{author}{Costello, E.}, \bibinfo{author}{Brunton, J.}, \bibinfo{author}{Brown, M.}, \bibinfo{author}{Daly, L.}, \bibinfo{year}{2018}.
\newblock \bibinfo{title}{In moocs we trust: Learner perceptions of mooc quality via trust and credibility.}
\newblock \bibinfo{journal}{International Journal of Emerging Technologies in Learning} \bibinfo{volume}{13}, \bibinfo{pages}{214–222}.
\newblock \DOIprefix\doi{https://doi.org/10.3991/ijet.v13i06.8447}.
\bibitem[{Cruzes and Dyba(2011)}]{diba2011recommended}
\bibinfo{author}{Cruzes, D.S.}, \bibinfo{author}{Dyba, T.}, \bibinfo{year}{2011}.
\newblock \bibinfo{title}{Recommended steps for thematic synthesis in software engineering}, in: \bibinfo{booktitle}{2011 international symposium on empirical software engineering and measurement}, \bibinfo{organization}{IEEE}. pp. \bibinfo{pages}{275--284}.
\newblock \DOIprefix\doi{10.1109/ESEM.2011.36}.
\bibitem[{Currall and Epstein(2003)}]{currall2003fragility}
\bibinfo{author}{Currall, S.C.}, \bibinfo{author}{Epstein, M.J.}, \bibinfo{year}{2003}.
\newblock \bibinfo{title}{The fragility of organizational trust:: Lessons from the rise and fall of enron}.
\newblock \bibinfo{journal}{Organizational Dynamics} \bibinfo{volume}{32}, \bibinfo{pages}{193--206}.
\newblock \DOIprefix\doi{https://doi.org/10.1016/S0090-2616(03)00018-4}.
\bibitem[{Deng et~al.(2018)Deng, Li and Huangfu}]{deng_2018_trust_ware_neural}
\bibinfo{author}{Deng, X.}, \bibinfo{author}{Li, H.}, \bibinfo{author}{Huangfu, F.}, \bibinfo{year}{2018}.
\newblock \bibinfo{title}{A trust-aware neural collaborative filtering for elearning recommendation}.
\newblock \bibinfo{journal}{Educational Sciences: Theory \& Practice} \bibinfo{volume}{18}, \bibinfo{pages}{2217--2234}.
\newblock \URLprefix \url{https://jestp.com/menuscript/index.php/estp/article/view/308/275}.
\bibitem[{Deutsch(1958)}]{deutsch1958trust}
\bibinfo{author}{Deutsch, M.}, \bibinfo{year}{1958}.
\newblock \bibinfo{title}{Trust and suspicion}.
\newblock \bibinfo{journal}{Journal of conflict resolution} \bibinfo{volume}{2}, \bibinfo{pages}{265--279}.
\newblock \DOIprefix\doi{https://doi.org/10.1177/002200275800200401}.
\bibitem[{Dorob{\u{a}}{\c{t}} et~al.(2019)Dorob{\u{a}}{\c{t}}, Corbea and Muntean}]{dorobuact_2019_integrating_student}
\bibinfo{author}{Dorob{\u{a}}{\c{t}}, I.}, \bibinfo{author}{Corbea, A.M.I.}, \bibinfo{author}{Muntean, M.}, \bibinfo{year}{2019}.
\newblock \bibinfo{title}{Integrating student trust in a conceptual model for assessing learning management system success in higher education: An empirical analysis}.
\newblock \bibinfo{journal}{IEEE Access} \bibinfo{volume}{7}, \bibinfo{pages}{69202--69214}.
\newblock \DOIprefix\doi{10.1109/ACCESS.2019.2919100}.
\bibitem[{Dwivedi and Bharadwaj(2013)}]{dwivedi_2013_effective_trust_ware}
\bibinfo{author}{Dwivedi, P.}, \bibinfo{author}{Bharadwaj, K.K.}, \bibinfo{year}{2013}.
\newblock \bibinfo{title}{Effective trust-aware e-learning recommender system based on learning styles and knowledge levels}.
\newblock \bibinfo{journal}{Journal of Educational Technology \& Society} \bibinfo{volume}{16}, \bibinfo{pages}{201--216}.
\newblock \DOIprefix\doi{https://www.jstor.org/stable/jeductechsoci.16.4.201}.
\bibitem[{Dwyer and Marsh(2016)}]{dwyer_2016_how_students_regard}
\bibinfo{author}{Dwyer, N.}, \bibinfo{author}{Marsh, S.}, \bibinfo{year}{2016}.
\newblock \bibinfo{title}{How students regard trust in an elearning context}, in: \bibinfo{booktitle}{2016 14th Annual Conference on Privacy, Security and Trust (PST)}, \bibinfo{organization}{IEEE}. pp. \bibinfo{pages}{682--685}.
\newblock \DOIprefix\doi{10.1109/PST.2016.7906956}.
\bibitem[{Elghomary and Bouzidi(2019)}]{elghomary_2019_dynamic_peer}
\bibinfo{author}{Elghomary, K.}, \bibinfo{author}{Bouzidi, D.}, \bibinfo{year}{2019}.
\newblock \bibinfo{title}{Dynamic peer recommendation system based on trust model for sustainable social tutoring in moocs}, in: \bibinfo{booktitle}{2019 1st International Conference on Smart Systems and Data Science (ICSSD)}, \bibinfo{organization}{IEEE}. pp. \bibinfo{pages}{1--9}.
\newblock \DOIprefix\doi{10.1109/ICSSD47982.2019.9003154}.
\bibitem[{Elghomary et~al.(2019)Elghomary, Bouzidi and Daoudi}]{elghomary_2019_a_comparative_analysis}
\bibinfo{author}{Elghomary, K.}, \bibinfo{author}{Bouzidi, D.}, \bibinfo{author}{Daoudi, N.}, \bibinfo{year}{2019}.
\newblock \bibinfo{title}{A comparative analysis of osn and siot trust models for a trust model adapted to moocs platforms}, in: \bibinfo{booktitle}{Proceedings of the 2nd International Conference on Networking, Information Systems \& Security}, pp. \bibinfo{pages}{1--8}.
\newblock \DOIprefix\doi{https://doi.org/10.1145/3320326.3320335}.
\bibitem[{Elghomary et~al.(2022)Elghomary, Bouzidi and Daoudi}]{Elghomary_Bouzidi_Daoudi_2022_design_of_smart_MOOC}
\bibinfo{author}{Elghomary, K.}, \bibinfo{author}{Bouzidi, D.}, \bibinfo{author}{Daoudi, N.}, \bibinfo{year}{2022}.
\newblock \bibinfo{title}{Design of a smart mooc trust model: Towards a dynamic peer recommendation to foster collaboration and learner’s engagement}.
\newblock \bibinfo{journal}{International Journal of Emerging Technologies in Learning (iJET)} \bibinfo{volume}{17}, \bibinfo{pages}{pp. 36–56}.
\newblock \DOIprefix\doi{10.3991/ijet.v17i05.27705}.
\bibitem[{Fachrunnisa et~al.(2010)Fachrunnisa, Hussain and Chang}]{fachrunnisa2010state}
\bibinfo{author}{Fachrunnisa, O.}, \bibinfo{author}{Hussain, F.}, \bibinfo{author}{Chang, E.}, \bibinfo{year}{2010}.
\newblock \bibinfo{title}{State of the art review for trust maintenance in organizations}, in: \bibinfo{booktitle}{2010 International Conference on Complex, Intelligent and Software Intensive Systems}, \bibinfo{organization}{IEEE}. pp. \bibinfo{pages}{574--580}.
\newblock \DOIprefix\doi{10.1109/CISIS.2010.194}.
\bibitem[{Feng and Xu(2020)}]{feng_2020_case_study}
\bibinfo{author}{Feng, C.}, \bibinfo{author}{Xu, Y.}, \bibinfo{year}{2020}.
\newblock \bibinfo{title}{Case study of collaborative learning in a massive open online course}, in: \bibinfo{booktitle}{2020 Ninth International Conference of Educational Innovation through Technology (EITT)}, \bibinfo{organization}{IEEE}. pp. \bibinfo{pages}{47--51}.
\newblock \DOIprefix\doi{10.1109/EITT50754.2020.00014}.
\bibitem[{Fujihara(2013)}]{fujihara_2013_toward_dependable_online}
\bibinfo{author}{Fujihara, Y.}, \bibinfo{year}{2013}.
\newblock \bibinfo{title}{Toward dependable online peer assessments a concept of the trust manegement on peer assessments}, in: \bibinfo{booktitle}{2013 IEEE Region 10 Humanitarian Technology Conference}, \bibinfo{organization}{IEEE}. pp. \bibinfo{pages}{110--111}.
\newblock \DOIprefix\doi{10.1109/R10-HTC.2013.6669024}.
\bibitem[{Ghosh et~al.(2001)Ghosh, Whipple and Bryan}]{ghosh2001student}
\bibinfo{author}{Ghosh, A.K.}, \bibinfo{author}{Whipple, T.W.}, \bibinfo{author}{Bryan, G.A.}, \bibinfo{year}{2001}.
\newblock \bibinfo{title}{Student trust and its antecedents in higher education}.
\newblock \bibinfo{journal}{The Journal of Higher Education} \bibinfo{volume}{72}, \bibinfo{pages}{322--340}.
\newblock \DOIprefix\doi{http://dx.doi.org/10.2307/2649334}.
\bibitem[{Golbeck(2005)}]{golbeck2005computing}
\bibinfo{author}{Golbeck, J.A.}, \bibinfo{year}{2005}.
\newblock \bibinfo{title}{Computing and applying trust in web-based social networks}.
\newblock \bibinfo{publisher}{University of Maryland, College Park}.
\newblock \URLprefix \url{https://hdl.handle.net/1903/2384}.
\bibitem[{Gutierrez et~al.(2015)Gutierrez, Osman, Roig and Sierra}]{gutierrez_2015_trust_based}
\bibinfo{author}{Gutierrez, P.}, \bibinfo{author}{Osman, N.}, \bibinfo{author}{Roig, C.}, \bibinfo{author}{Sierra, C.}, \bibinfo{year}{2015}.
\newblock \bibinfo{title}{Trust-based community assessment}.
\newblock \bibinfo{journal}{Pattern Recognition Letters} \bibinfo{volume}{67}, \bibinfo{pages}{49--58}.
\newblock \DOIprefix\doi{https://doi.org/10.1016/j.patrec.2015.06.016}.
\bibitem[{Horvat et~al.(2013)Horvat, Krsmanovic, Dobrota and \v{C}udanov}]{horvat2013students}
\bibinfo{author}{Horvat, A.}, \bibinfo{author}{Krsmanovic, M.}, \bibinfo{author}{Dobrota, M.}, \bibinfo{author}{\v{C}udanov, M.}, \bibinfo{year}{2013}.
\newblock \bibinfo{title}{Students` trust in distance learning: Changes in satisfaction and significance}.
\newblock \bibinfo{journal}{Management: Journal for Theory and Practice Management} \bibinfo{volume}{18}, \bibinfo{pages}{47--54}.
\newblock \DOIprefix\doi{https://scindeks.ceon.rs/article.aspx?artid=0354-86351369047H}.
\bibitem[{Jones and Shah(2021)}]{jones2021tangled}
\bibinfo{author}{Jones, S.}, \bibinfo{author}{Shah, P.}, \bibinfo{year}{2021}.
\newblock \bibinfo{title}{The tangled ties of trust: A social network perspective on interpersonal trust}, in: \bibinfo{booktitle}{Understanding Trust in Organizations}. \bibinfo{publisher}{Taylor and Francis Inc.}, pp. \bibinfo{pages}{205--232}.
\newblock \DOIprefix\doi{https://doi.org/10.4324/9780429449185}.
\bibitem[{K{\"a}hk{\"o}nen et~al.(2021)K{\"a}hk{\"o}nen, Blomqvist, Gillespie and Vanhala}]{kahkonen2021employee}
\bibinfo{author}{K{\"a}hk{\"o}nen, T.}, \bibinfo{author}{Blomqvist, K.}, \bibinfo{author}{Gillespie, N.}, \bibinfo{author}{Vanhala, M.}, \bibinfo{year}{2021}.
\newblock \bibinfo{title}{Employee trust repair: A systematic review of 20 years of empirical research and future research directions}.
\newblock \bibinfo{journal}{Journal of Business Research} \bibinfo{volume}{130}, \bibinfo{pages}{98--109}.
\newblock \DOIprefix\doi{https://doi.org/10.1016/j.jbusres.2021.03.019}.
\bibitem[{Khan et~al.(2023)Khan, Cram, Wang, Tran, Cavaleri and Rahman}]{khan2023modelling}
\bibinfo{author}{Khan, E.A.}, \bibinfo{author}{Cram, A.}, \bibinfo{author}{Wang, X.}, \bibinfo{author}{Tran, K.}, \bibinfo{author}{Cavaleri, M.}, \bibinfo{author}{Rahman, M.J.}, \bibinfo{year}{2023}.
\newblock \bibinfo{title}{Modelling the impact of online learning quality on students' satisfaction, trust and loyalty}.
\newblock \bibinfo{journal}{International Journal of Educational Management} \bibinfo{volume}{37}, \bibinfo{pages}{281--299}.
\newblock \DOIprefix\doi{https://doi.org/10.1108/IJEM-02-2022-0066}.
\bibitem[{Kitchenham and Charters(2007)}]{kitchenham2007guidelines}
\bibinfo{author}{Kitchenham, B.}, \bibinfo{author}{Charters, S.}, \bibinfo{year}{2007}.
\newblock \bibinfo{title}{Guidelines for performing systematic literature reviews in software engineering}.
\newblock \bibinfo{type}{Technical Report}. Keele University, Keele, UK.
\newblock \URLprefix \url{https://legacyfileshare.elsevier.com/promis_misc/525444systematicreviewsguide.pdf}.
\bibitem[{Kitchenham et~al.(2011)Kitchenham, Budgen and Brereton}]{kitchenham2011using}
\bibinfo{author}{Kitchenham, B.A.}, \bibinfo{author}{Budgen, D.}, \bibinfo{author}{Brereton, O.P.}, \bibinfo{year}{2011}.
\newblock \bibinfo{title}{Using mapping studies as the basis for further research--a participant-observer case study}.
\newblock \bibinfo{journal}{Information and Software Technology} \bibinfo{volume}{53}, \bibinfo{pages}{638--651}.
\newblock \DOIprefix\doi{https://doi.org/10.1016/j.infsof.2010.12.011}.
\bibitem[{Kitchenham et~al.(2015)Kitchenham, Budgen and Brereton}]{kitchenham2015evidence}
\bibinfo{author}{Kitchenham, B.A.}, \bibinfo{author}{Budgen, D.}, \bibinfo{author}{Brereton, P.}, \bibinfo{year}{2015}.
\newblock \bibinfo{title}{Evidence-based software engineering and systematic reviews}. volume~\bibinfo{volume}{4}.
\newblock \bibinfo{publisher}{CRC press}.
\newblock \DOIprefix\doi{https://doi.org/10.1201/b19467}.
\bibitem[{Knack and Zak(2003)}]{knack2003building}
\bibinfo{author}{Knack, S.}, \bibinfo{author}{Zak, P.J.}, \bibinfo{year}{2003}.
\newblock \bibinfo{title}{Building trust: public policy, interpersonal trust, and economic development}.
\newblock \bibinfo{journal}{Supreme court economic review} \bibinfo{volume}{10}, \bibinfo{pages}{91--107}.
\newblock \DOIprefix\doi{https://www.journals.uchicago.edu/doi/10.1086/scer.10.1147139}.
\bibitem[{Koufaris and Hampton-Sosa(2004)}]{koufaris2004development}
\bibinfo{author}{Koufaris, M.}, \bibinfo{author}{Hampton-Sosa, W.}, \bibinfo{year}{2004}.
\newblock \bibinfo{title}{The development of initial trust in an online company by new customers}.
\newblock \bibinfo{journal}{Information \& management} \bibinfo{volume}{41}, \bibinfo{pages}{377--397}.
\newblock \DOIprefix\doi{https://doi.org/10.1016/j.im.2003.08.004}.
\bibitem[{Kunthi et~al.(2018)Kunthi, Wahyuni, Al-Hafidz and Sensuse}]{kunthi_2018_exploring_antecedent}
\bibinfo{author}{Kunthi, R.}, \bibinfo{author}{Wahyuni, R.}, \bibinfo{author}{Al-Hafidz, M.U.}, \bibinfo{author}{Sensuse, D.I.}, \bibinfo{year}{2018}.
\newblock \bibinfo{title}{Exploring antecedent factors toward knowledge sharing intention in e-learning}, in: \bibinfo{booktitle}{2018 IEEE Symposium on Computer Applications \& Industrial Electronics (ISCAIE)}, \bibinfo{organization}{IEEE}. pp. \bibinfo{pages}{108--113}.
\newblock \DOIprefix\doi{10.1109/ISCAIE.2018.8405453}.
\bibitem[{Kutsyuruba and Walker(2015)}]{kutsyuruba2015lifecycle}
\bibinfo{author}{Kutsyuruba, B.}, \bibinfo{author}{Walker, K.}, \bibinfo{year}{2015}.
\newblock \bibinfo{title}{The lifecycle of trust in educational leadership: An ecological perspective}.
\newblock \bibinfo{journal}{International Journal of Leadership in Education} \bibinfo{volume}{18}, \bibinfo{pages}{106--121}.
\newblock \DOIprefix\doi{https://doi.org/10.1080/13603124.2014.915061}.
\bibitem[{Laifa et~al.(2015)Laifa, Giglou, Akhrouf and Maamri}]{laifa_2015_trust_and_forgiveness}
\bibinfo{author}{Laifa, M.}, \bibinfo{author}{Giglou, R.I.}, \bibinfo{author}{Akhrouf, S.}, \bibinfo{author}{Maamri, R.}, \bibinfo{year}{2015}.
\newblock \bibinfo{title}{Trust and forgiveness in virtual teams: A study in algerian e-learning context}, in: \bibinfo{booktitle}{2015 International Conference on Interactive Mobile Communication Technologies and Learning (IMCL)}, \bibinfo{organization}{IEEE}. pp. \bibinfo{pages}{131--135}.
\newblock \DOIprefix\doi{http://dx.doi.org/10.1109/IMCTL.2015.7359571}.
\bibitem[{Li et~al.(2015)Li, Bao, Zheng and Huang}]{li_2015_social_analytics}
\bibinfo{author}{Li, Y.}, \bibinfo{author}{Bao, H.}, \bibinfo{author}{Zheng, Y.}, \bibinfo{author}{Huang, Z.}, \bibinfo{year}{2015}.
\newblock \bibinfo{title}{Social analytics framework to boost recommendation in online learning communities}, in: \bibinfo{booktitle}{2015 IEEE 15th International Conference on Advanced Learning Technologies}, \bibinfo{organization}{IEEE}. pp. \bibinfo{pages}{405--406}.
\newblock \DOIprefix\doi{10.1109/ICALT.2015.100}.
\bibitem[{Lohikoski et~al.(2014)Lohikoski, Muhos and H{\"a}rk{\"o}nen}]{lohikoski2014virtual}
\bibinfo{author}{Lohikoski, P.}, \bibinfo{author}{Muhos, M.}, \bibinfo{author}{H{\"a}rk{\"o}nen, J.}, \bibinfo{year}{2014}.
\newblock \bibinfo{title}{Virtual collaboration competence requirements for entrepreneurship education in sparsely populated areas}, in: \bibinfo{booktitle}{ICEL-2014 Hosted by The Federico Santa Mar{\'\i}a Technical University Valparaiso Chile}, pp. \bibinfo{pages}{109--117}.
\bibitem[{Mahamud et~al.(2021)Mahamud, Fam, Saleh, Kamarudin and Wahjono}]{mahamud2021predicting}
\bibinfo{author}{Mahamud, S.}, \bibinfo{author}{Fam, S.F.}, \bibinfo{author}{Saleh, H.}, \bibinfo{author}{Kamarudin, M.F.}, \bibinfo{author}{Wahjono, S.I.}, \bibinfo{year}{2021}.
\newblock \bibinfo{title}{Predicting google classroom acceptance and use in stem education: Extended utaut2 approach}, in: \bibinfo{booktitle}{2021 2nd SEA-STEM International Conference (SEA-STEM)}, \bibinfo{organization}{IEEE}. pp. \bibinfo{pages}{155--159}.
\newblock \DOIprefix\doi{10.1109/SEA-STEM53614.2021.9668096}.
\bibitem[{Mao and Shen(2016)}]{mao_2016_web_of_credit}
\bibinfo{author}{Mao, Y.}, \bibinfo{author}{Shen, H.}, \bibinfo{year}{2016}.
\newblock \bibinfo{title}{Web of credit: Adaptive personalized trust network inference from online rating data}.
\newblock \bibinfo{journal}{IEEE Transactions on Computational Social Systems} \bibinfo{volume}{3}, \bibinfo{pages}{176--189}.
\newblock \DOIprefix\doi{10.1109/TCSS.2016.2639016}.
\bibitem[{Mayer et~al.(1995)Mayer, Davis and Schoorman}]{mayer1995integrative}
\bibinfo{author}{Mayer, R.C.}, \bibinfo{author}{Davis, J.H.}, \bibinfo{author}{Schoorman, F.D.}, \bibinfo{year}{1995}.
\newblock \bibinfo{title}{An integrative model of organizational trust}.
\newblock \bibinfo{journal}{Academy of management review} \bibinfo{volume}{20}, \bibinfo{pages}{709--734}.
\newblock \DOIprefix\doi{https://doi.org/10.2307/258792}.
\bibitem[{Mayfield and Valenti(2022)}]{mayfield2022team}
\bibinfo{author}{Mayfield, C.O.}, \bibinfo{author}{Valenti, A.}, \bibinfo{year}{2022}.
\newblock \bibinfo{title}{Team satisfaction, identity, and trust: a comparison of face-to-face and virtual student teams}.
\newblock \bibinfo{journal}{Active Learning in Higher Education} , \bibinfo{pages}{213--226}\DOIprefix\doi{https://doi.org/10.1177/14697874221118861}.
\bibitem[{McEvily et~al.(2021)McEvily, Zaheer and Soda}]{mcevily2021network}
\bibinfo{author}{McEvily, B.}, \bibinfo{author}{Zaheer, A.}, \bibinfo{author}{Soda, G.}, \bibinfo{year}{2021}.
\newblock \bibinfo{title}{Network trust}, in: \bibinfo{booktitle}{Understanding trust in organizations}. \bibinfo{publisher}{Taylor \& Francis}, pp. \bibinfo{pages}{179--204}.
\newblock \DOIprefix\doi{https://doi.org/10.4324/9780429449185}.
\bibitem[{Medema et~al.(2014)Medema, Wals and Adamowski}]{medema_2014_multi_loop}
\bibinfo{author}{Medema, W.}, \bibinfo{author}{Wals, A.}, \bibinfo{author}{Adamowski, J.}, \bibinfo{year}{2014}.
\newblock \bibinfo{title}{Multi-loop social learning for sustainable land and water governance: Towards a research agenda on the potential of virtual learning platforms}.
\newblock \bibinfo{journal}{NJAS-Wageningen Journal of Life Sciences} \bibinfo{volume}{69}, \bibinfo{pages}{23--38}.
\newblock \DOIprefix\doi{https://doi.org/10.1016/j.njas.2014.03.003}.
\bibitem[{Miguel et~al.(2014a)Miguel, Caball{\'e}, Xhafa and Prieto}]{miguel_2014_security_in_online}
\bibinfo{author}{Miguel, J.}, \bibinfo{author}{Caball{\'e}, S.}, \bibinfo{author}{Xhafa, F.}, \bibinfo{author}{Prieto, J.}, \bibinfo{year}{2014}a.
\newblock \bibinfo{title}{Security in online learning assessment towards an effective trustworthiness approach to support e-learning teams}, in: \bibinfo{booktitle}{2014 IEEE 28th International Conference on Advanced Information Networking and Applications}, \bibinfo{organization}{IEEE}. pp. \bibinfo{pages}{123--130}.
\newblock \DOIprefix\doi{10.1109/AINA.2014.106}.
\bibitem[{Miguel et~al.(2014b)Miguel, Caball{\'e}, Xhafa, Prieto and Barolli}]{miguel2014collective}
\bibinfo{author}{Miguel, J.}, \bibinfo{author}{Caball{\'e}, S.}, \bibinfo{author}{Xhafa, F.}, \bibinfo{author}{Prieto, J.}, \bibinfo{author}{Barolli, L.}, \bibinfo{year}{2014}b.
\newblock \bibinfo{title}{A collective intelligence approach for building student's trustworthiness profile in online learning}, in: \bibinfo{booktitle}{2014 Ninth International Conference on P2P, Parallel, Grid, Cloud and Internet Computing}, \bibinfo{organization}{IEEE}. pp. \bibinfo{pages}{46--53}.
\newblock \DOIprefix\doi{10.1109/3PGCIC.2014.132}.
\bibitem[{Miguel et~al.(2014c)Miguel, Caball{\'e}, Xhafa, Prieto and Barolli}]{miguel_2014_predicting_trustworthiness}
\bibinfo{author}{Miguel, J.}, \bibinfo{author}{Caball{\'e}, S.}, \bibinfo{author}{Xhafa, F.}, \bibinfo{author}{Prieto, J.}, \bibinfo{author}{Barolli, L.}, \bibinfo{year}{2014}c.
\newblock \bibinfo{title}{Predicting trustworthiness behavior to enhance security in on-line assessment}, in: \bibinfo{booktitle}{2014 International Conference on Intelligent Networking and Collaborative Systems}, \bibinfo{organization}{IEEE}. pp. \bibinfo{pages}{342--349}.
\newblock \DOIprefix\doi{10.1109/INCoS.2014.19}.
\bibitem[{Morrison et~al.(2012)Morrison, Cegielski and Rainer}]{morrison_2012_trust_Avatar}
\bibinfo{author}{Morrison, R.}, \bibinfo{author}{Cegielski, C.G.}, \bibinfo{author}{Rainer, R.K.}, \bibinfo{year}{2012}.
\newblock \bibinfo{title}{Trust, avatars, and electronic communications: Implications for e-learning}.
\newblock \bibinfo{journal}{Journal of Computer Information Systems} \bibinfo{volume}{53}, \bibinfo{pages}{80--89}.
\newblock \DOIprefix\doi{https://www.tandfonline.com/doi/abs/10.1080/08874417.2012.11645600}.
\bibitem[{Napole{\~a}o et~al.(2021)Napole{\~a}o, Felizardo, de~Souza, Petrillo, Hall{\'e}, Vijaykumar and Nakagawa}]{napoleao2021establishing}
\bibinfo{author}{Napole{\~a}o, B.M.}, \bibinfo{author}{Felizardo, K.R.}, \bibinfo{author}{de~Souza, {\'E}.F.}, \bibinfo{author}{Petrillo, F.}, \bibinfo{author}{Hall{\'e}, S.}, \bibinfo{author}{Vijaykumar, N.L.}, \bibinfo{author}{Nakagawa, E.Y.}, \bibinfo{year}{2021}.
\newblock \bibinfo{title}{Establishing a search string to detect secondary studies in software engineering}, in: \bibinfo{booktitle}{2021 47th Euromicro Conference on Software Engineering and Advanced Applications (SEAA)}, \bibinfo{organization}{IEEE}. pp. \bibinfo{pages}{9--16}.
\newblock \DOIprefix\doi{10.1109/SEAA53835.2021.00010}.
\bibitem[{Nigam et~al.(2021)Nigam, Pasricha, Singh and Churi}]{nigam_2021_a_systematic_review}
\bibinfo{author}{Nigam, A.}, \bibinfo{author}{Pasricha, R.}, \bibinfo{author}{Singh, T.}, \bibinfo{author}{Churi, P.}, \bibinfo{year}{2021}.
\newblock \bibinfo{title}{A systematic review on ai-based proctoring systems: Past, present and future}.
\newblock \bibinfo{journal}{Education and Information Technologies} \bibinfo{volume}{26}, \bibinfo{pages}{6421--6445}.
\newblock \DOIprefix\doi{https://doi.org/10.1007/s10639-021-10597-x}.
\bibitem[{Noh et~al.(2021)Noh, Jang and Jeon}]{noh_2021_the_influence_of}
\bibinfo{author}{Noh, J.S.}, \bibinfo{author}{Jang, H.Y.}, \bibinfo{author}{Jeon, S.S.}, \bibinfo{year}{2021}.
\newblock \bibinfo{title}{The influence of e-wom information characteristics on learning trust and e-wom intention among online learning}.
\newblock \bibinfo{journal}{Journal of Logistics, Informatics and Service Science} \bibinfo{volume}{8}, \bibinfo{pages}{134--150}.
\newblock \DOIprefix\doi{http://dx.doi.org/10.33168/JSMS.2023.0218}.
\bibitem[{Petersen et~al.(2015)Petersen, Vakkalanka and Kuzniarz}]{petersen2015guidelines}
\bibinfo{author}{Petersen, K.}, \bibinfo{author}{Vakkalanka, S.}, \bibinfo{author}{Kuzniarz, L.}, \bibinfo{year}{2015}.
\newblock \bibinfo{title}{Guidelines for conducting systematic mapping studies in software engineering: An update}.
\newblock \bibinfo{journal}{Information and software technology} \bibinfo{volume}{64}, \bibinfo{pages}{1--18}.
\newblock \DOIprefix\doi{https://doi.org/10.1016/j.infsof.2015.03.007}.
\bibitem[{Pham et~al.(2020)Pham, Vu and Tran}]{pham_2020_the_role_of}
\bibinfo{author}{Pham, C.}, \bibinfo{author}{Vu, N.}, \bibinfo{author}{Tran, G.}, \bibinfo{year}{2020}.
\newblock \bibinfo{title}{The role of e-learning service quality and e-trust on e-loyalty}.
\newblock \bibinfo{journal}{Management Science Letters} \bibinfo{volume}{10}, \bibinfo{pages}{2741--2750}.
\newblock \DOIprefix\doi{http://dx.doi.org/10.5267/j.msl.2020.4.036}.
\bibitem[{Robinson(1996)}]{robinson1996trust}
\bibinfo{author}{Robinson, S.L.}, \bibinfo{year}{1996}.
\newblock \bibinfo{title}{Trust and breach of the psychological contract}.
\newblock \bibinfo{journal}{Administrative science quarterly} , \bibinfo{pages}{574--599}\DOIprefix\doi{https://doi.org/10.2307/2393868}.
\bibitem[{Rotter(1967)}]{rotter1967new}
\bibinfo{author}{Rotter, J.B.}, \bibinfo{year}{1967}.
\newblock \bibinfo{title}{A new scale for the measurement of interpersonal trust.}
\newblock \bibinfo{journal}{Journal of personality} , \bibinfo{pages}{651--665}\DOIprefix\doi{https://psycnet.apa.org/doi/10.1111/j.1467-6494.1967.tb01454.x}.
\bibitem[{Rousseau et~al.(1998)Rousseau, Sitkin, Burt and Camerer}]{rousseau1998not}
\bibinfo{author}{Rousseau, D.M.}, \bibinfo{author}{Sitkin, S.B.}, \bibinfo{author}{Burt, R.S.}, \bibinfo{author}{Camerer, C.}, \bibinfo{year}{1998}.
\newblock \bibinfo{title}{Not so different after all: A cross-discipline view of trust}.
\newblock \bibinfo{journal}{Academy of management review} \bibinfo{volume}{23}, \bibinfo{pages}{393--404}.
\newblock \DOIprefix\doi{http://dx.doi.org/10.5465/AMR.1998.926617}.
\bibitem[{Schaubroeck et~al.(2013)Schaubroeck, Peng and Hannah}]{schaubroeck2013developing}
\bibinfo{author}{Schaubroeck, J.M.}, \bibinfo{author}{Peng, A.C.}, \bibinfo{author}{Hannah, S.T.}, \bibinfo{year}{2013}.
\newblock \bibinfo{title}{Developing trust with peers and leaders: Impacts on organizational identification and performance during entry}.
\newblock \bibinfo{journal}{Academy of Management Journal} \bibinfo{volume}{56}, \bibinfo{pages}{1148--1168}.
\newblock \DOIprefix\doi{https://doi.org/10.5465/amj.2011.0358}.
\bibitem[{Sedighehm and Ainin(2018)}]{sedighehm_2018_effect}
\bibinfo{author}{Sedighehm, M.}, \bibinfo{author}{Ainin, S.}, \bibinfo{year}{2018}.
\newblock \bibinfo{title}{Effect of trust and perceived reciprocal benefit on students knowledge sharing via fb and academic performance}.
\newblock \bibinfo{journal}{The Electronic Journal of Knowledge Management} \bibinfo{volume}{16}, \bibinfo{pages}{23--35}.
\newblock \DOIprefix\doi{https://researchportal.murdoch.edu.au/esploro/outputs/journalArticle/Effect-of-trust-and-perceived-reciprocal/991005541304007891}.
\bibitem[{Snyman(2021)}]{snyman2021framework}
\bibinfo{author}{Snyman, A.M.}, \bibinfo{year}{2021}.
\newblock \bibinfo{title}{A framework for staff retention in the higher education environment: effects of the psychological contract, organisational justice and trust}.
\newblock \URLprefix \url{https://uir.unisa.ac.za/bitstream/handle/10500/27910/thesis_snyman_am.pdf?sequence=1\&isAllowed=y}.
\bibitem[{Surahman and Wang(2022)}]{surahman_2022_academic_dishonesty}
\bibinfo{author}{Surahman, E.}, \bibinfo{author}{Wang, T.H.}, \bibinfo{year}{2022}.
\newblock \bibinfo{title}{Academic dishonesty and trustworthy assessment in online learning: a systematic literature review}.
\newblock \bibinfo{journal}{Journal of Computer Assisted Learning} \bibinfo{volume}{38}, \bibinfo{pages}{1535--1553}.
\newblock \DOIprefix\doi{https://doi.org/10.1111/jcal.12708}.
\bibitem[{Sztompka(1999)}]{sztompka1999trust}
\bibinfo{author}{Sztompka, P.}, \bibinfo{year}{1999}.
\newblock \bibinfo{title}{Trust: A sociological theory}.
\newblock \bibinfo{publisher}{Cambridge university press}.
\newblock \DOIprefix\doi{10.1017/S0266267102221136}.
\bibitem[{Taghizadeh~Kerman et~al.(2022)Taghizadeh~Kerman, Banihashem, Noroozi and Biemans}]{taghizadeh2022effects}
\bibinfo{author}{Taghizadeh~Kerman, N.}, \bibinfo{author}{Banihashem, S.K.}, \bibinfo{author}{Noroozi, O.}, \bibinfo{author}{Biemans, H.J.}, \bibinfo{year}{2022}.
\newblock \bibinfo{title}{The effects of students’ perceived usefulness and trustworthiness of peer feedback on learning satisfaction in online learning environments}, in: \bibinfo{booktitle}{HEAd 2022-8th International Conference on Higher Education Advances}, \bibinfo{publisher}{Editorial Universitat Polit{\`e}cnica de Val{\`e}ncia}. pp. \bibinfo{pages}{263--271}.
\newblock \DOIprefix\doi{http://dx.doi.org/10.4995/HEAd22.2022.14445}.
\bibitem[{Tan et~al.(2012)Tan, Li, Tang, Wang and Hu}]{A2}
\bibinfo{author}{Tan, W.}, \bibinfo{author}{Li, J.}, \bibinfo{author}{Tang, A.}, \bibinfo{author}{Wang, T.}, \bibinfo{author}{Hu, X.}, \bibinfo{year}{2012}.
\newblock \bibinfo{title}{Trust evaluation model based on user trust cloud and user capability in e-learning service}, in: \bibinfo{booktitle}{Communications and Information Processing: International Conference, ICCIP 2012 Aveiro, Portugal, March 7-11, 2012 Revised Selected Papers, Part I}, \bibinfo{organization}{Springer}. pp. \bibinfo{pages}{583--590}.
\newblock \DOIprefix\doi{http://dx.doi.org/10.1007/978-3-642-31965-5_68}.
\bibitem[{Trainer and Redmiles(2018)}]{trainer2018bridging}
\bibinfo{author}{Trainer, E.H.}, \bibinfo{author}{Redmiles, D.F.}, \bibinfo{year}{2018}.
\newblock \bibinfo{title}{Bridging the gap between awareness and trust in globally distributed software teams}.
\newblock \bibinfo{journal}{Journal of Systems and Software} \bibinfo{volume}{144}, \bibinfo{pages}{328--341}.
\newblock \DOIprefix\doi{https://doi.org/10.1016/j.jss.2018.06.028}.
\bibitem[{Tsankova et~al.(2017)Tsankova, Marinov, Durcheva and Varbanova}]{tsankova_2017_synergy_effect}
\bibinfo{author}{Tsankova, R.}, \bibinfo{author}{Marinov, O.}, \bibinfo{author}{Durcheva, M.}, \bibinfo{author}{Varbanova, E.}, \bibinfo{year}{2017}.
\newblock \bibinfo{title}{Synergy effect of the tesla project in management of engineering higher education}, in: \bibinfo{booktitle}{Proceedings of the 9th International Conference on Management of Digital EcoSystems}, pp. \bibinfo{pages}{259--264}.
\newblock \DOIprefix\doi{https://doi.org/10.1145/3167020.3167059}.
\bibitem[{Tseng et~al.(2019)Tseng, Yeh and Tang}]{tseng_2019_a_close_look}
\bibinfo{author}{Tseng, H.}, \bibinfo{author}{Yeh, H.T.}, \bibinfo{author}{Tang, Y.}, \bibinfo{year}{2019}.
\newblock \bibinfo{title}{A close look at trust among team members in online learning communities}.
\newblock \bibinfo{journal}{International Journal of Distance Education Technologies (IJDET)} \bibinfo{volume}{17}, \bibinfo{pages}{52--65}.
\newblock \DOIprefix\doi{10.4018/IJDET.2019010104}.
\bibitem[{Udhayakumar et~al.(2021)Udhayakumar, Uma~Nandhini and Chandrasekaran}]{udhayakumar_2021_trusted_cooperative}
\bibinfo{author}{Udhayakumar, S.}, \bibinfo{author}{Uma~Nandhini, D.}, \bibinfo{author}{Chandrasekaran, S.}, \bibinfo{year}{2021}.
\newblock \bibinfo{title}{Trusted cooperative e-learning service deployment model in multi-cloud environment}, in: \bibinfo{booktitle}{Inventive Communication and Computational Technologies: Proceedings of ICICCT 2020}, \bibinfo{organization}{Springer}. pp. \bibinfo{pages}{527--535}.
\newblock \DOIprefix\doi{https://doi.org/10.1007/978-981-15-7345-3_45}.
\bibitem[{Valtolina et~al.(2022)Valtolina, Matamoros, Musiu, Epifania and Villa}]{valtolina_2022_extended_utaut}
\bibinfo{author}{Valtolina, S.}, \bibinfo{author}{Matamoros, R.A.}, \bibinfo{author}{Musiu, E.}, \bibinfo{author}{Epifania, F.}, \bibinfo{author}{Villa, M.}, \bibinfo{year}{2022}.
\newblock \bibinfo{title}{Extended utaut model to analyze the acceptance of virtual assistant’s recommendations using interactive visualisations}, in: \bibinfo{booktitle}{Proceedings of the 2022 International Conference on Advanced Visual Interfaces}, pp. \bibinfo{pages}{1--5}.
\newblock \DOIprefix\doi{https://doi.org/10.1145/3531073.3531129}.
\bibitem[{Van~Solingen et~al.(2002)Van~Solingen, Basili, Caldiera and Rombach}]{solingengoal}
\bibinfo{author}{Van~Solingen, R.}, \bibinfo{author}{Basili, V.}, \bibinfo{author}{Caldiera, G.}, \bibinfo{author}{Rombach, H.D.}, \bibinfo{year}{2002}.
\newblock \bibinfo{title}{Goal question metric (gqm) approach}.
\newblock \bibinfo{journal}{Encyclopedia of software engineering} \DOIprefix\doi{https://doi.org/10.1002/0471028959.sof142}.
\bibitem[{Wang and Chen(2012)}]{wang_2012_emotions_and_pair}
\bibinfo{author}{Wang, M.j.}, \bibinfo{author}{Chen, H.C.}, \bibinfo{year}{2012}.
\newblock \bibinfo{title}{Emotions and pair trust in asynchronous hospitality cultural exchange for students in taiwan and hong kong}.
\newblock \bibinfo{journal}{Turkish Online Journal of Educational Technology} \bibinfo{volume}{11}, \bibinfo{pages}{119 – 131}.
\newblock \DOIprefix\doi{http://www.tojet.net/articles/v11i4/11411.pdf}.
\bibitem[{Wang(2014)}]{wang_2014_building_student_trust}
\bibinfo{author}{Wang, Y.D.}, \bibinfo{year}{2014}.
\newblock \bibinfo{title}{Building student trust in online learning environments}.
\newblock \bibinfo{journal}{Distance Education} \bibinfo{volume}{35}, \bibinfo{pages}{345--359}.
\bibitem[{Wang and Emurian(2005)}]{wang2005overview}
\bibinfo{author}{Wang, Y.D.}, \bibinfo{author}{Emurian, H.H.}, \bibinfo{year}{2005}.
\newblock \bibinfo{title}{An overview of online trust: Concepts, elements, and implications}.
\newblock \bibinfo{journal}{Computers in human behavior} \bibinfo{volume}{21}, \bibinfo{pages}{105--125}.
\bibitem[{Widjaja et~al.(2016)Widjaja, Chen and Hiele}]{widjaja_2016_the_effect_of}
\bibinfo{author}{Widjaja, A.E.}, \bibinfo{author}{Chen, J.V.}, \bibinfo{author}{Hiele, T.M.}, \bibinfo{year}{2016}.
\newblock \bibinfo{title}{The effect of online participation in online learning course for studying trust in information and communication technologies}.
\newblock \bibinfo{journal}{International Journal of Cyber Behavior, Psychology and Learning (IJCBPL)} \bibinfo{volume}{6}, \bibinfo{pages}{79--93}.
\newblock \DOIprefix\doi{10.4018/IJCBPL.2016070106}.
\bibitem[{Widjaja et~al.(2020)Widjaja, Santoso, Fernando, Condrobimo et~al.}]{widjaja_2020_improving_the_quality}
\bibinfo{author}{Widjaja, H.A.E.}, \bibinfo{author}{Santoso, S.W.}, \bibinfo{author}{Fernando, E.}, \bibinfo{author}{Condrobimo, A.R.}, et~al., \bibinfo{year}{2020}.
\newblock \bibinfo{title}{Improving the quality of learning management system (lms) based on student perspectives using utaut2 and trust model}, in: \bibinfo{booktitle}{2020 4th International Conference on Informatics and Computational Sciences (ICICoS)}, \bibinfo{organization}{IEEE}. pp. \bibinfo{pages}{1--6}.
\newblock \DOIprefix\doi{10.1109/ICICoS51170.2020.9298985}.
\bibitem[{Wohlin et~al.(2012)Wohlin, Runeson, H{\"o}st, Ohlsson, Regnell and Wessl{\'e}n}]{wohlin2012experimentation}
\bibinfo{author}{Wohlin, C.}, \bibinfo{author}{Runeson, P.}, \bibinfo{author}{H{\"o}st, M.}, \bibinfo{author}{Ohlsson, M.C.}, \bibinfo{author}{Regnell, B.}, \bibinfo{author}{Wessl{\'e}n, A.}, \bibinfo{year}{2012}.
\newblock \bibinfo{title}{Experimentation in software engineering}.
\newblock \bibinfo{publisher}{Springer Science \& Business Media}.
\newblock \DOIprefix\doi{10.1007/978-3-642-29044-2}.
\bibitem[{Wongse-ek et~al.(2013)Wongse-ek, Wills and Gilbert}]{wongse_2013_towards_a_trust_model}
\bibinfo{author}{Wongse-ek, W.}, \bibinfo{author}{Wills, G.B.}, \bibinfo{author}{Gilbert, L.}, \bibinfo{year}{2013}.
\newblock \bibinfo{title}{Towards a trust model in e-learning: Antecedents of a student's trust.}
\newblock \bibinfo{journal}{International Association for Development of the Information Society} \DOIprefix\doi{https://files.eric.ed.gov/fulltext/ED562293.pdf}.
\bibitem[{Wu and Zhang(2015)}]{wu_2015_trust_evalueation}
\bibinfo{author}{Wu, B.}, \bibinfo{author}{Zhang, C.}, \bibinfo{year}{2015}.
\newblock \bibinfo{title}{Trust evaluation for inter-organization knowledge sharing via the e-learning community}.
\newblock \bibinfo{journal}{The Electronic Library} \bibinfo{volume}{33}, \bibinfo{pages}{400--416}.
\newblock \DOIprefix\doi{https://doi.org/10.1108/EL-08-2013-0140}.
\bibitem[{Younas et~al.(2021)Younas, Faisal, Habib, Ashraf and Ahmad}]{younas_2021_role_of_design}
\bibinfo{author}{Younas, A.}, \bibinfo{author}{Faisal, C.N.}, \bibinfo{author}{Habib, M.A.}, \bibinfo{author}{Ashraf, R.}, \bibinfo{author}{Ahmad, M.}, \bibinfo{year}{2021}.
\newblock \bibinfo{title}{Role of design attributes to determine the intention to use online learning via cognitive beliefs}.
\newblock \bibinfo{journal}{Ieee Access} \bibinfo{volume}{9}, \bibinfo{pages}{94181--94202}.
\newblock \DOIprefix\doi{10.1109/ACCESS.2021.3093348}.
\bibitem[{Zhang et~al.(2022)Zhang, Li, Seng, Chen and Chen}]{zhang_2022_a_novel_precise}
\bibinfo{author}{Zhang, X.}, \bibinfo{author}{Li, M.}, \bibinfo{author}{Seng, D.}, \bibinfo{author}{Chen, X.}, \bibinfo{author}{Chen, X.}, \bibinfo{year}{2022}.
\newblock \bibinfo{title}{A novel precise personalized learning recommendation model regularized with trust and influence}.
\newblock \bibinfo{journal}{Scientific Programming} \bibinfo{volume}{2022}, \bibinfo{pages}{1--15}.
\newblock \DOIprefix\doi{https://doi.org/10.1155/2022/8479423}.

\end{thebibliography}


\begin{thebibliography}{00}


\bibitem[Lamport(1994)]{lamport94}
  Leslie Lamport,
  \textit{\LaTeX: a document preparation system},
  Addison Wesley, Massachusetts,
  2nd edition,
  1994.

\end{thebibliography}

\end{document}